\documentclass[prc,superscriptaddress
,nofootinbib,
tightenlines]{revtex4}

\usepackage{epsfig}
\usepackage{slashed}
\usepackage{amssymb}
\usepackage{amsmath}
\usepackage{color}
\usepackage{hyperref}

\newcommand{\beq}{\begin{equation}}
\newcommand{\eeq}{\end{equation}}
\newcommand{\beqa}{\begin{eqnarray}}
\newcommand{\eeqa}{\end{eqnarray}}

\newcommand{\ben}{\begin{displaymath}}
\newcommand{\een}{\end{displaymath}}
\newcommand{\be}{\begin{equation}}
\newcommand{\ee}{\end{equation}}
\newcommand{\bea}{\begin{eqnarray}}
\newcommand{\eea}{\end{eqnarray}}
\newcommand{\fet}[1]{\mbox{\boldmath $#1$}}

\begin{document}

\title{Scrutiny of the new class of three-nucleon forces}
\author{E.~Epelbaum}
\affiliation
{Institut f\" ur Theoretische Physik II, Fakult\" at f\" ur Physik und Astronomie,
Ruhr-Universit\" at Bochum, D-44780 Bochum, Germany}
\author{A.~M.~Gasparyan}
\affiliation
{Institut f\" ur Theoretische Physik II, Fakult\" at f\" ur Physik und Astronomie,
Ruhr-Universit\" at Bochum, D-44780 Bochum, Germany}
\author{J.~Gegelia}
\affiliation
{Institut f\" ur Theoretische Physik II, Fakult\" at f\" ur Physik und Astronomie,
Ruhr-Universit\" at Bochum, D-44780 Bochum, Germany}  
\affiliation{Tbilisi State  University,  0186 Tbilisi,
 Georgia}
\author{D.~Hog}
\affiliation
{Institut f\" ur Theoretische Physik II, Fakult\" at f\" ur Physik und Astronomie,
Ruhr-Universit\" at Bochum, D-44780 Bochum, Germany}
\author{H.~Krebs}
\affiliation
{Institut f\" ur Theoretische Physik II, Fakult\" at f\" ur Physik und Astronomie,
Ruhr-Universit\" at Bochum, D-44780 Bochum, Germany}

\begin{abstract}
In a recent publication, Cirigliano {\it et al.} [Phys. Rev. Lett. 135,
022501 (2025)] argue that three-nucleon forces (3NFs) involving short-range operators that couple
two pions with two nucleons are enhanced beyond what is expected in
chiral effective field theory based on naive dimensional
analysis. Here, we scrutinize the arguments and conclusions of that paper
by taking into account renormalization 
scheme dependence of the corresponding low-energy constants. We gain
further insights into the expected impact of these 3NFs by comparing
them with contributions of similar type, induced by pion-exchange
diagrams at lower orders in the chiral expansion. We also estimate
the impact of these 3NFs on properties of nuclear matter.
After removal of scheme-dependent
short-distance components in pion loops, the 3NFs considered by
Cirigliano {\it et al.}~are shown to yield reasonably small contributions to
the equation of state of neutron and symmetric nuclear matter in agreement
with expectations based on Weinberg's power counting.  
\end{abstract}


\maketitle

\section{Introduction}

Recent years have seen impressive progress in {\it ab initio} description of
atomic nuclei and properties of nuclear matter \cite{Hergert:2020bxy}. At the same time,
chiral effective field theory (EFT)
\cite{Epelbaum:2008ga,Machleidt:2011zz} has been pushed in the two-nucleon (NN) sector 
to fifth expansion order (N$^4$LO) and even beyond,
leading to the development of high-precision NN
potentials capable of a statistically perfect description of mutually
compatible neutron-proton and proton-proton scattering data below pion
production threshold \cite{Reinert:2017usi,Reinert:2020mcu}. In view of these developments, a detailed
quantitative understanding of 3NFs is becoming increasingly more
urgent and pressing \cite{Kalantar-Nayestanaki:2011rzs,Hammer:2012id,Endo:2024cbz}.

\begin{figure}[tb]
	\includegraphics[width=0.9\textwidth]{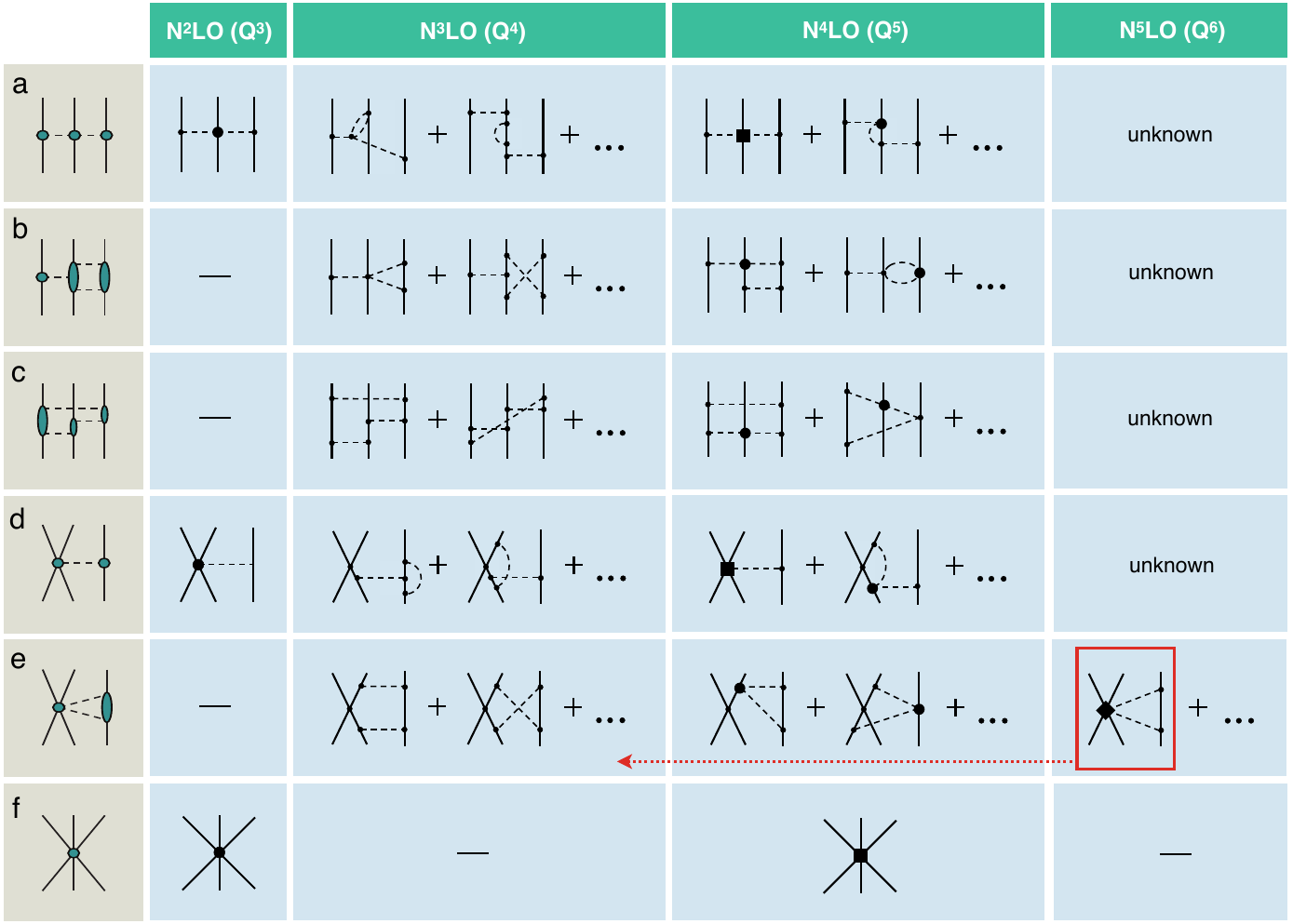}
	\caption{Diagrams contributing to the 3NF in chiral EFT based on
          Weinberg's power counting with pions and nucleons as the
          only explicit degrees of freedom. Up to N$^4$LO, all possible
          3NF topologies are shown in the leftmost column and include the
          two-pion exchange (a),  two-pion-one-pion exchange (b), ring
        (c), one-pion-exchange-contact (d), two-pion-exchange-contact
        (e) and purely contact (f) diagrams. Dashed and solid lines
        denote pions and nucleons, respectively. Solid dots, filled
        circles, filled diamonds and filled squares denote the
        vertices from the effective chiral Lagrangian of dimension
        $\Delta_i = 0$, $1$, $2$ and $3$, respectively. The dominant N$^2$LO
        contributions are derived in Refs.~\cite{vanKolck:1994yi,Epelbaum:2002vt}. The 
        expressions for the N$^3$LO contributions, calculated using
        dimensional regularization, can be found in Refs.~\cite{Ishikawa:2007zz,Bernard:2007sp,Bernard:2011zr},
        while the N$^4$LO corrections of types (a), (b) and (c) have
        been worked out in Refs.~\cite{Krebs:2012yv,Krebs:2013kha}. The subleading contact 3NF
        of type (f) is discussed in Ref.~\cite{Girlanda:2011fh}, while the N$^4$LO
        contributions of the type (d) and (e) have not yet been worked
        out. The diagram in the
        last column is considered by Cirigliano {\it et al.}~\cite{Cirigliano:2024ocg} and
        argued to be enhanced beyond NDA as explained in the text. }
	\label{fig1}
\end{figure}

The current status of the derivation of 3NFs
using the Weinberg power counting of chiral EFT with pions and
nucleons as the only active degrees of freedom is summarized in
Fig.~\ref{fig1}. The low-momentum scaling $Q^\nu$ of an $N$-nucleon
connected irreducible diagram, where $Q = \{M_\pi/\Lambda_b, \, |\vec p \, |/\Lambda_b\}$
is the expansion parameter with $M_\pi$, $\vec p$ and
$\Lambda_b$ referring to the pion mass, typical nucleon momentum and
the breakdown scale, respectively, can be obtained from naive
dimensional analysis (NDA) \cite{Weinberg:1990rz,Weinberg:1991um,Epelbaum:2007us}: $\nu = -4 + 2 N + 2L + \sum_i V_i
\Delta_i$. Here, $L$ denotes the number of loops and the sum goes
over all vertices appearing in a diagram. Furthermore, $V_i$ is the
number of vertices of type $i$ while the vertex dimension $\Delta_i$
is defined as $\Delta_i = d_i + n_i/2 - 2$, with $d_i$ and $n_i$ being the number
of derivatives and/or $M_\pi$-insertions and nucleon field operators,
respectively. The dominant 3NFs at next-to-next-to-leading order (N$^2$LO) in the
chiral expansion are nowadays routinely taken into account in  {\it
  ab initio} calculations and have been found to provide important
contributions to three-nucleon (3N) scattering observables, properties
of nuclei, nuclear reactions as well as the equation of state (EoS) of
nuclear matter, see Refs.~\cite{LENPIC:2022cyu,Somasundaram:2024ykk,Yang:2025mhg,Alp:2025wjn} for selected recent
applications. The extension of these studies to include 3NF
contributions beyond N$^2$LO is complicated by issues related to
regularization, since mixing dimensional regularization in the derivation of
the 3NF with cutoff regulators in the Schr\"odinger/Faddeev
equation violates chiral symmetry \cite{Epelbaum:2019kcf}. A
rigorous regularization method based on the chiral gradient flow,
which preserves the chiral and gauge symmetries, has been proposed in
Refs.~\cite{Krebs:2023ljo,Krebs:2023gge} and is currently being
applied to rederive the 3NFs beyond N$^2$LO.  

In a recent paper \cite{Cirigliano:2024ocg}, Cirigliano {\it et al.} consider a particular
contribution to the two-pion-exchange-contact 3NF of type (e) at
order $Q^6$ (N$^5$LO), stemming from a triangle one-loop diagram
with a  quark-mass-dependent $\pi\pi$NN vertex, see the rightmost
graph in Fig.~\ref{fig1}. They argue that such 3NF contributions are
enhanced by a factor of $Q^{-2}$ compared to the estimation based on NDA and thus
contribute already at N$^3$LO. The line of arguments and main
results of Ref.~\cite{Cirigliano:2024ocg} can be summarized as follows: 
\begin{itemize}
  \item
The authors start with reiterating the findings of Ref.~\cite{Kaplan:1996xu} regarding
the renormalization group (RG) behavior of the low-energy constant (LEC)
$D_2$ that accompanies a quark-mass dependent derivative-less NN contact interaction. They
then conclude that {\it ``$D_2$ is needed at LO in
approaches to Chiral EFT [...] that aim to ensure regulator
independence''}. In other words, the NN contact interaction $\propto
D_2 M_\pi^2$ is argued to be enhanced by two inverse powers of the
expansion parameter and thus claimed to
contribute at LO ($Q^0$) rather than NLO ($Q^2$) as expected in
Weinberg's power counting  based on NDA.  For typical cutoff values employed in chiral EFT
calculations, the authors of Ref.~\cite{Cirigliano:2024ocg}  estimate $|D_2 | \lesssim
10$~fm$^4$, but they consider a
smaller variation of $|D_2 | \lesssim 4$~fm$^4$ in their numerical estimations. 
\item
  An enhanced value of $D_2$ necessarily implies the enhanced 
  vertices with four nucleon fields and any even 
  number of pions, whose strength is determined by $D_2$ by
  virtue of spontaneously broken chiral symmetry. Accordingly, 
   the two-pion-exchange-contact 3NF $\propto D_2$
  in the rightmost column of Fig.~\ref{fig1} is argued to contribute at order
  $Q^4$ (N$^3$LO) rather than $Q^6$ (N$^5$LO). 
\item
  Motivated by these arguments, the authors of Ref.~\cite{Cirigliano:2024ocg} derive the expressions for the
  type-(e) 3NF $\propto D_2$ and $F_2$, where $F_2$
  denotes the LEC of the quark-mass independent $\pi\pi NN$
  vertex with two derivatives, using dimensional regularization to
  calculate pion loops. Assuming $|D_2 |, \, |F_2 | \lesssim
  4$~fm$^4$, they find very large effects in nuclear matter, which seems to support the need to
  promote these 3NFs to a lower order. 
  \end{itemize}
Clearly, these conclusions, if valid, put pressure on the mainstream
applications of chiral EFT to nuclear systems relying on the
NDA-based hierarchy of nuclear forces as reviewed in
Refs.~\cite{Epelbaum:2008ga,Machleidt:2011zz,Epelbaum:2019kcf}. In this paper, we critically address the arguments put
forward in Ref.~\cite{Cirigliano:2024ocg} and take a closer look at
the convergence pattern of the chiral expansion for 3NFs, focusing
especially on the type-(e) topology. We also 
estimate the contributions of the considered 3NFs to the EoS of nuclear
matter. 

Our paper is organized as follows. In sec.~\ref{sec:RG}, we discuss
the RG equation and the resulting scaling of the LEC $D_2$ for different
choices of renormalization conditions. We argue that the
enhanced size of $D_2$ assumed in
the analysis of Ref.~\cite{Cirigliano:2024ocg}, $D_2 \sim \mathcal{O}(Q^{-2})$, corresponds to the
choice of renormalization conditions employed by Kaplan, Savage and
Wise (KSW) \cite{Kaplan:1998tg}, while the LEC $D_2$ is expected to scale according
to NDA, i.e.~$D_2 \sim \mathcal{O}(1)$, in the Weinberg scheme. We
also estimate the size of $D_2$ by supplementing the RG equation
with numerical values of the leading-order (LO) NN contact
interactions taken from the state-of-the-art chiral NN potentials of
Refs.~\cite{Reinert:2017usi,Epelbaum:2022cyo}. To gain further insights into the expected size of the
considered 3NFs and, more generally, into the convergence pattern of
chiral EFT for nuclear potentials, the novel 3NFs
are compared in
sec.~\ref{sec:ChEFT3NF} with the parameter-free N$^3$LO and N$^4$LO contributions
of type (e) induced by three-pion exchange diagrams of type
(b). Motivated by the similarity between the 3NFs
$\propto D_2, \, F_2$ and the subleading
$2\pi$-exchange NN potential, we take a closer look at the convergence pattern of chiral EFT in
the NN sector  in
sec.~\ref{sec:ChEFT2NF}. We argue/recall that removing (scheme-dependent)
short-range contributions from the $2\pi$-exchange potential, calculated using dimensional
regularization, is essential for uncovering the full predictive power
of chiral EFT in the NN sector
\cite{Epelbaum:2003gr,Epelbaum:2014efa,Reinert:2017usi,Epelbaum:2024gfg}. Using
regularized $t$-channel dispersion relations to remove short-range components from 
pion loops and employing the values of the LECs $D_2$ and $F_2$ consistent with
Weinberg's power counting, we estimate the impact of various 3NF
contributions to the EoS of pure neutron and symmetric nuclear matter
in sec.~\ref{sec:EoS}. The main results of our study are
summarized in sec.~\ref{sec:Summary}.

\section{Renormalization group arguments and the scaling of $D_2$}
\label{sec:RG}

The quest for a consistent renormalization in the few-nucleon sector
of chiral EFT has attracted much attention during last decades. It seems
undisputed that RG arguments play an important role in this context,
yet a universally accepted understanding of what this role is actually
supposed to be is still lacking. The dependence of renormalized
couplings on renormalization points in quantum field theory (QFT) is
controlled by the Gell-Mann and Low RG equations \cite{Gell-Mann:1954yli}.
An alternative view on renormalization is offered by the
Wilsonian approach \cite{Wilson:1973jj}, which addresses RG trajectories
in the space of bare coupling constants as functions of cutoff parameter(s).
Clearly, the full expressions of physical quantities do not depend on the
applied renormalization scheme, i.e., the exact scattering amplitudes
remain constant along Wilsonian RG trajectories. On the other
hand, the whole utility of the RG method lies in its ability to
reorganize perturbation series in such a way that perturbative
contributions remain small. The aim here is to reduce the magnitude of
higher-order corrections, thereby improving reliability of
perturbative calculations \cite{Collins:1984xc}.

An essential feature of RG applied to perturbative calculations in QFT
is the {\it renormalization point(s)/cutoff-dependence} of approximate 
perturbative expressions for physical quantities. By exploiting
scale-dependence  of finite sums of perturbative series, one aims at choosing
such values of renormalization points/cutoff, which lead to an
optimal convergence of perturbative series for observables of
interest. Notice that while a renormalization scheme yielding a meaningful
perturbative expansion for the amplitude may not even exist depending
on the problem under consideration, it is always possible
to spoil the convergence of a perturbative series by choosing
inappropriate renormalization schemes.  

Clearly, the general features of the RG mentioned above also apply to
chiral EFT and its application to nuclear systems. The relative
importance of terms appearing in the effective Lagrangian, the power
counting, depends on the choice of renormalization scale(s) or
subtraction points in calculations utilizing
the standard QFT renormalization by subtracting divergences.
Alternatively, if one uses the Wilsonian approach, power
counting rules depend on the choice of cutoff parameter(s).

The
application of chiral EFT to the two-nucleon system is complicated by the
fine-tuned nature of the NN S-wave scattering. In the near-threshold region,
the on-shell scattering amplitude $T$ can be parametrized in terms of the
effective range expansion (ERE), 
\beq
T \; =\;  - \frac{4 \pi}{m} \frac{1}{p \cot \delta - i p} \; =  \;  -
\frac{4 \pi}{m} \frac{1}{(-1/a + 1/2 r p^2 + \ldots) - i p} \,,
\eeq
where $m$ is the nucleon mass, $p$ is the on-shell momentum, $\delta$ is the phase shift,
while $a$ and $r$ refer to the scattering length and effective range,
respectively. The fine-tuned nature of the problem at hand manifests
itself in the small numerical values of the first coefficient in the
ERE, $|a^{-1}| \ll M_\pi$, in both the spin-singlet
($^1$S$_0$) and triplet ($^3$S$_1$) partial waves.
For $|a^{-1}| \ll p \ll M_\pi$, the
scaling of the amplitude changes from $T \sim \mathcal{O}(1)$ expected
by NDA to $T \sim \mathcal{O}(Q^{-1})$\footnote{For
  pionless EFT, the expansion parameter $Q$ is defined as $Q =
  p/\Lambda_b$ with $\Lambda_b \sim M_\pi$.} characteristic to S-wave systems close to the unitary
limit and signals the non-perturbative nature of the NN interaction at
very low energy.

The above-mentioned features must be taken into
account when formulating EFTs for NN scattering. Below,
we outline different EFT formulations in the NN sector
and discuss implications for the scaling of the LEC $D_2$ relevant for
this study. 

\subsection{KSW scheme}
\label{sec:KSW}

\begin{figure}[tb]
	\includegraphics[width=0.85\textwidth]{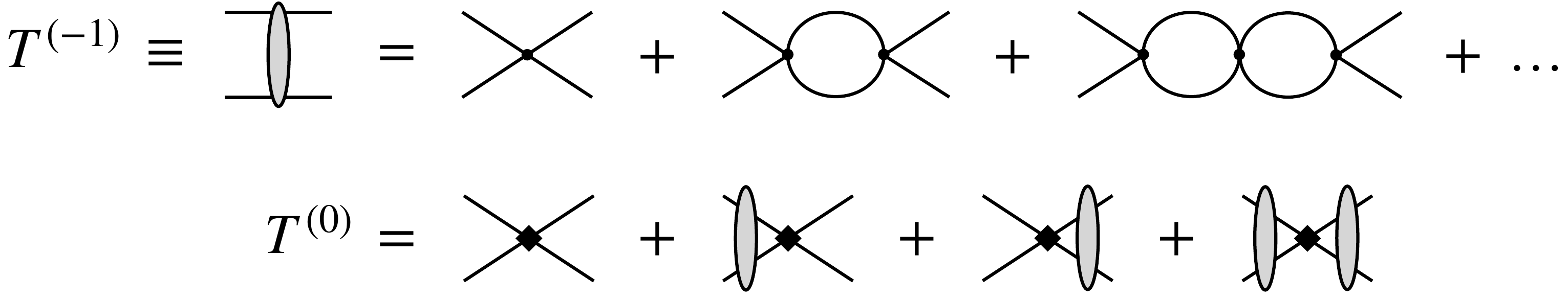}
	\caption{The leading and subleading contributions to the
          S-wave NN
          scattering amplitude in pionless EFT. For notation see Fig.~\ref{fig1}.}
	\label{fig2}
\end{figure}

The most frequently used formulation of pionless EFT was
proposed by Kaplan, Savage and Wise \cite{Kaplan:1998tg}, see also \cite{vanKolck:1997ut,vanKolck:1998bw} for a
closely related work. In this scheme,
renormalization conditions are fixed by choosing all
renormalization scale(s) or, equivalently, momentum subtraction
points \cite{Mehen:1998zz} of the order of the soft scale, i.e.~$\mu \sim p \ll M_\pi \sim
\Lambda_b$. 
 For this choice of renormalization conditions,
matching pionless EFT to the fine-tuned NN scattering amplitude requires the
values of renormalized LECs $C_0 (\mu)$, $C_2 (\mu)$, $\ldots$ to be 
enhanced beyond NDA: $C_0 (\mu) \sim \mathcal{O} (Q^{-1})$, $C_2 (\mu)
\sim \mathcal{O} (Q^{-2})$, $\ldots$. Here, the
subscript $n$ of a NN LEC $C_n$ denotes the power of
momenta at the corresponding vertex. After renormalization, all diagrams in the upper row of
Fig.~\ref{fig2} contribute at order $Q^{-1}$ and must be resummed to
generate the LO amplitude $T^{(-1)}$. 
Corrections beyond LO are generated by subleading interactions
using distorted-wave perturbation theory, as exemplified in
Fig.~\ref{fig2} for the case of the subleading amplitude $T^{(0)}$. 

The KSW scheme can be straightforwardly generalized to include
pions. Since the enhancement of the amplitude beyond NDA is achieved
by using enhanced values of the NN contact interactions, $C_n (\mu ) \sim \mathcal{O} (Q^{-n/2-1})$,
pion-exchange contributions are subleading in this scheme and thus treated perturbatively
\cite{Kaplan:1998tg}. The resulting framework with perturbative pions
allows one to {\it explicitly}
renormalize the NN scattering amplitude, but has been shown to lack
convergence in low spin-triplet partial waves for energies outside of the
applicability range of pionless EFT \cite{Cohen:1998jr,Fleming:1999ee}, see,
however, Ref.~\cite{Kaplan:2019znu} for a more optimistic conclusion.

\subsection{(Quasi-) Weinberg scheme}
\label{sec:Weinberg}

Weinberg's power
counting scheme with renormalized LECs scaling according to NDA
can be realized by choosing the subtraction scale $\mu$ of linearly
divergent integrals of the order of the hard scale in the problem,
$\mu \sim \Lambda_b$, and all other renormalization/subtraction scales 
of the order of the soft scale \cite{Epelbaum:2017byx}. In pionless EFT with
this choice of renormalization conditions, the LO contribution to the NN
amplitude is still given by diagrams shown in the upper row of
Fig.~\ref{fig2}, which are all linearly divergent and thus scale as
$\sim \mathcal{O} (1)$. Fine tuning is implemented by tuning the value
of $C_0
(\mu)$ such that the amplitude stemming from the
resummed LO bubble diagrams has a pole around $| p | \approx 0$, i.e., the
resummed LO amplitude scales as $\sim \mathcal{O} (Q^{-1})$. The
subleading contribution to the amplitude in this scheme is generated
by the last diagram in the second row of Fig.~\ref{fig2}, while the
first, second and third graphs contribute at orders $Q^2$,  $Q^1$ and
$Q^1$, respectively.

The scaling of renormalized LECs $C_0$ and $C_2$
for both choices of renormalization conditions can be easily verified
by calculating all diagrams in Fig.~\ref{fig2} and matching the resulting
amplitude to the first two terms in the ERE. Setting, for the sake of
simplicity, the subtraction scale of cubic divergences
to zero one finds 
\beq
\label{C0C2}
C_0 (\mu) \; =\; \frac{4 \pi}{m}\, \frac{1}{a^{-1} - \mu}, \qquad
C_2 (\mu) \; =\; \frac{\pi}{m}\, \frac{r}{(a^{-1} - \mu)^2},
\eeq
which yields $C_0 (\mu) \sim \mathcal{O} (Q^{-1})$,  $C_2 (\mu) \sim
\mathcal{O} (Q^{-2})$ in the KSW scheme with $\mu \sim p \ll
\Lambda_b$ while $C_0 (\mu) \sim C_2 (\mu) \sim \mathcal{O} (1)$ in
the Weinberg scheme with $\mu \sim \Lambda_b$. Notice that despite the
fact that the two approaches lead to different sets of diagrams
contributing to the amplitude beyond LO, they provide self-consistent
and, in fact, equivalent formulations of pionless EFT for NN
scattering. The choices of renormalization conditions to describe even
stronger fine-tuned P-wave systems in the framework of halo-EFT are discussed in detail
in Ref.~\cite{Epelbaum:2021sns}. 

While equivalent in pionless EFT, the KSW and Weinberg choices of
renormalization conditions lead to different formulations of chiral
EFT. Since the enhancement of two-nucleon-reducible diagrams
in the Weinberg scheme is achieved by choosing $\mu \sim \Lambda_b$,
it equally applies to both the iterated  LO contact interaction
$C_0$ and the one-pion ($1\pi$) exchange potential 
\beq
\label{1pi}
V_{1\pi} \; = \; - \frac{g_A^2}{4F_\pi^2} \fet \tau_1 \cdot \fet
\tau_2 \frac{\vec \sigma_1 \cdot \vec q \, \vec \sigma_2 \cdot \vec
  q}{q^2 + M_\pi^2}\,,
\eeq
which feature a similar ultraviolet behavior (except for spin-singlet
channels as will be discussed in sec.~\ref{sec:ScalingD2}). Here, $F_\pi$  and $g_A$ are the pion
decay and nucleon axial-vector couplings, respectively, while $\vec
\sigma_i$ ($\fet \tau_i$) denote the Pauli spin (isospin) matrices of
nucleon $i$. Accordingly, the $1\pi$-exchange potential must, in general, be
treated nonperturbatively in the Weinberg scheme. 

\subsection{Finite-cutoff chiral EFT}
\label{sec:FiniteCutoff}

Clearly, an {\it explicit} renormalization of the amplitude with
nonperturbatively treated $1\pi$-exchange potential is a formidable
task. A more practical and frequently used approach, which is also well suited for
applications beyond the two-nucleon system, is based on a finite-cutoff
formulation of chiral EFT along the lines of Ref.~\cite{Lepage:1997cs}, see
Ref.~\cite{Epelbaum:2019kcf} for a detailed description and
Refs.~\cite{Wesolowski:2021cni,LENPIC:2022cyu,Somasundaram:2024ykk,Chambers-Wall:2024fha,Gennari:2024sbn,Alp:2025wjn,Yang:2025mhg}
for selected recent applications. While not required by power
counting, nuclear potentials are
typically treated nonperturbatively
in the Schr\"odinger/Lippmann-Schwinger equation within this scheme. Since not all counterterms needed for absorbing UV
divergences in multi-loop few-nucleon-reducible diagrams can be
included in practice, the momentum cutoff $\Lambda$ has
to be chosen of the order of the breakdown scale, $\Lambda \sim
\Lambda_b$ \cite{Lepage:1997cs,Epelbaum:2009sd}. Instead of performing {\it explicit} renormalization
as outlined in sections \ref{sec:KSW} and \ref{sec:Weinberg},
scattering amplitudes/observables are {\it implicitly} renormalized by tuning
bare LECs $C_0 (\Lambda )$, $C_2 (\Lambda )$, $\ldots$, which are
expected to be of natural size in accordance with NDA, to low-energy
experimental data. Accordingly, power counting is less transparent and not manifest
in this scheme,\footnote{In the pionless case, finite-cutoff EFT
can be solved analytically in the two-nucleon sector and is equivalent to the formulations described in sections
  \ref{sec:KSW}  and \ref{sec:Weinberg}. It is conjectured in
  Ref.~\cite{Epelbaum:2017byx} that for natural values of $C_n (\Lambda)$, 
  the finite-cutoff chiral EFT is equivalent to
  the formulation described in
  sec.~\ref{sec:Weinberg}.} but it can, in principle,
be verified {\it a posteriori} using (Bayesian) statistical methods. In particular,
a regular convergence pattern of finite-cutoff chiral EFT in the
formulation of Ref.~\cite{Reinert:2017usi} was confirmed  in
Ref.~\cite{Millican:2025sdp} for the cutoff choices of $\Lambda
= 450$ and $500$~MeV. We further emphasize that renormalizability of
a finite-cutoff chiral EFT in the NN sector was rigorously proven at
next-to-leading order (NLO) in Refs.~\cite{Gasparyan:2021edy,Gasparyan:2023rtj}.

\subsection{RG-invariant chiral EFT}
\label{sec:RGInvariant}

A different view on renormalization is taken within the so-called
RG-invariant formulation of chiral EFT by demanding the existence of
the $\Lambda \to \infty$ limit for the scattering amplitude at each
expansion order \cite{Hammer:2019poc,vanKolck:2020llt}. In this
scheme, the LO contributions to the NN potential, including the
$1\pi$-exchange, are treated non-perturbatively while corrections
beyond LO are included in perturbation theory. Power
counting is determined not by analyzing the contributions of
renormalized $C_n$ to the amplitude, but rather inferred from stabilizing
its $\Lambda \gg \Lambda_b$ behavior. Our critical view of this approach
is explained in detail in Refs.~\cite{Epelbaum:2018zli,Tews:2022yfb},
with the key conceptual issue being the EFT validity
of resummed UV-divergent, partially renormalized perturbative series
in the regime with $\Lambda
\gg \Lambda_b$.

\subsection{RG scaling of $D_2$}
\label{sec:ScalingD2}

\begin{figure}[tb]
	\includegraphics[width=\textwidth]{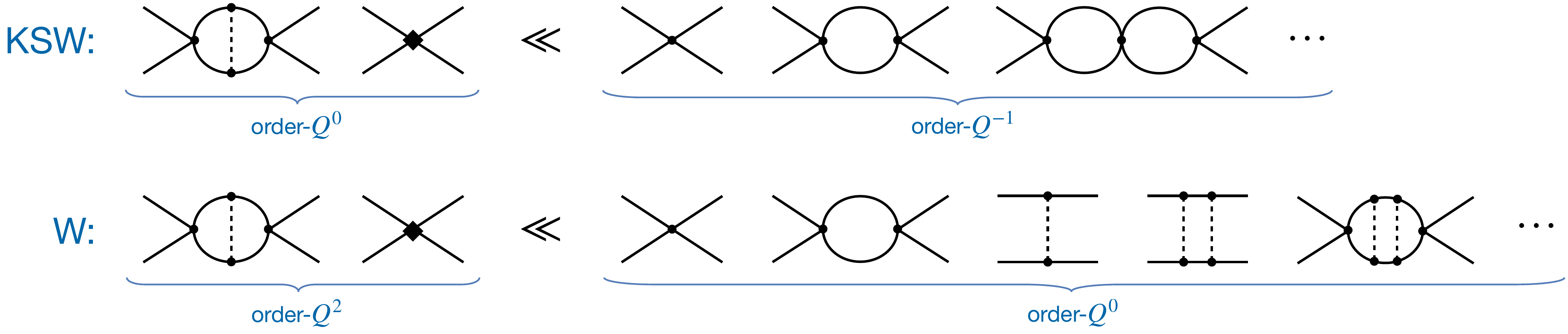}
	\caption{Scaling of various diagrams for the KSW and Weinberg
          (W) choices of renormalization conditions specified in
          sections \ref{sec:KSW} and \ref{sec:Weinberg}
          are shown in the upper and lower rows,
          respectively. The scaling of diagrams involving two
          potential pions is
          shown for spin-triplet NN channels, while the corresponding
          contributions in spin-singlet channels are
          suppressed. Notice further that for the near-threshold kinematics,
         the contributions of the first two
          (subleading) diagrams in the Weinberg scheme  get
          enhanced by the factor of $Q^{-2}$ after dressing them with the LO
          amplitude stemming from resummed diagrams on the right-hand
          side of the inequality. For remaining notation see Fig.~\ref{fig1}. }
	\label{fig3}
\end{figure}

Having outlined different chiral EFT formulations for
nuclear systems, we are now in the position to take a closer look at the
RG behavior of the LEC $D_2$ and discuss implications for the
hierarchy of 3NFs. The key observation made in Ref.~\cite{Kaplan:1996xu} and
reiterated by Cirigliano {\it et al.}~\cite{Cirigliano:2024ocg} concerns the
$1\pi$-exchange diagram dressed by the LO contact interactions $C_0$,
see the first graph in each row of Fig.~\ref{fig3}.  
This diagram is logarithmically divergent. A removal of the
logarithmic divergence requires the inclusion of the NN counterterm
$\propto D_2 M_\pi^2$, see the second diagram  in each row of Fig.~\ref{fig3}.  
In Ref.~\cite{Kaplan:1996xu}, this observation was argued to provide 
evidence of inconsistency of Weinberg's power counting, which
assigns the $1\pi$-exchange to LO while treating the operator $D_2
M_\pi^2$ as a subleading correction. It is instructive to scrutinize this argument
using chiral EFT formulations with manifest power counting described in sections \ref{sec:KSW} and
\ref{sec:Weinberg}. The
renormalized contribution of the first diagram in Fig.~\ref{fig3} in
the $^1$S$_0$ channel has the form
\beq
\label{DressedOPEP}
\frac{g_A^2}{4 (4 \pi F_\pi)^2} m^2 M_\pi^2 C_0^2 (\mu) \Big[ \ln \nu
+ f \big(p / M_\pi ) \Big]\,,
\eeq
where $f(x)$ is some function, whose explicit form is not relevant for
our discussion, and $\nu$ is the
scale/subtraction point of the logarithmic divergence. In the KSW
scheme with $C_0 (\mu )|_{\mu \sim M_\pi} \sim \mathcal{O} (Q^{-1})$, this contribution
scales as $\sim \mathcal{O} (1)$ and thus appears at NLO
together with the operator $D_2(\mu, \nu) M_\pi^2$, which ensures
$\nu$-independence of the amplitude at this order. In the Weinberg
scheme of sec.~\ref{sec:Weinberg} with $C_0 (\mu )|_{\mu \sim
  \Lambda_b} \sim \mathcal{O} (1)$, this diagram contributes at
order $Q^2$ as does the $D_2(\mu, \nu) M_\pi^2$-operator with
$D_2(\mu, \nu) |_{\mu \sim  \Lambda_b} \sim \mathcal{O} (1)$. It should,
however, be understood that the contributions of the $1\pi$-exchange and $D_2(\mu, \nu)
M_\pi^2$ term, dressed with
the full LO amplitude, are enhanced by the factor of $Q^{-2}$ and 
appear at subleading order $\mathcal{O} (1)$ like in the KSW scheme. 
These considerations demonstrate that the widespread claim about the
inconsistency of the Weinberg power counting in the $^1S_0$  partial wave
\cite{Kaplan:1998tg,Beane:2001bc} based on the appearance of a
logarithmic divergence in the dressed $1\pi$-exchange diagram is not
correct\footnote{We note, however, that Weinberg's original way of formally justifying
  the nonperturbative resummation in the LO amplitude by 
  counting the nucleon mass according to $m M_\pi \sim \Lambda_b^2$ \cite{Weinberg:1990rz,Weinberg:1991um},
  i.e.~$m \sim \mathcal{O} (Q^{-1})$, would indeed formally shift the
  expression in Eq.~(\ref{DressedOPEP}) to order $\mathcal{O} (1)$.}.

We also note in passing that Weinberg's choice of renormalization
conditions naturally explains the observed small/large impact of the
$1\pi$-exchange potential in the $^1$S$_0$/$^3$S$_1$ partial waves.  
To see this we write the $1\pi$-exchange in terms of the 
tensor operator $T_{12}(\vec q) : = \vec \sigma_1 \cdot \vec q \, \vec \sigma_2
\cdot \vec q - 1/3 \, q^2\, \vec \sigma_1 \cdot \vec \sigma_2$ as
\beq
\label{OPEP}
V_{1\pi} \; = \; - \frac{g_A^2}{4F_\pi^2} \fet \tau_1 \cdot \fet
\tau_2 \bigg(\frac{T_{12}(\vec q\,)}{q^2 + M_\pi^2} \; -\;  \frac{1}{3}  \vec
\sigma_1 \cdot \vec \sigma_2 \frac{M_\pi^2}{q^2 + M_\pi^2} \; + \;  \frac{1}{3}  \vec
\sigma_1 \cdot \vec \sigma_2\bigg) \,.
\eeq
The constant term can be absorbed into the LO contact interaction
$C_0$ (as was implicitly assumed in the above discussion).  The singular
tensor part of the $1\pi$-exchange $\propto T_{12}(\vec q\, )$ does not contribute in
spin-singlet channels. The remaining term $\propto M_\pi^2$ is
non-singular, and its iterations cannot generate enhanced linear
divergences. In contrast, in the $^3$S$_1$-$^3$D$_1$ channel, diagrams involving
multiple $1\pi$-exchange insertions like the last two graphs
in Fig.~\ref{fig3} are enhanced for $\mu \sim \Lambda_b$ and have to be
resummed. 

Even if not required from power counting, the $1\pi$-exchange can still
be included nonperturbatively in the $^1$S$_0$ channel, as done, e.g.,
in the framework outlined in sec.~\ref{sec:FiniteCutoff}. Contrary to
the philosophy of the RG-invariant chiral EFT approach described in sec.~\ref{sec:RGInvariant}
and to the argument put forward by Cirigliano {\it et al.}~\cite{Cirigliano:2024ocg}, the
inclusion of the $1\pi$-exchange potential at LO does {\it not} necessitate
the need to promote the $D_2 M_\pi^2$-operator to LO for properly
chosen renormalization conditions. In particular, assuming that $C_0
(\mu)|_{\mu \sim \Lambda_b}$ is of natural size, the $\nu$-dependent contribution in Eq.~(\ref{DressedOPEP})
would only become comparable to LO terms for very large values of $\nu$ of the order of $\nu \sim e^{\Lambda_b^2/M_\pi^2}$ (expressed in units of $M_\pi$).  In contrast, choosing $\nu $ of the order of any mass scale in
the problem like $M_\pi$, $\Lambda_b$ or $m$ leads to scale-dependent
contributions beyond the LO accuracy. 

The appearance of residual (i.e., of a higher order) renormalization scale dependence
is, in fact, a common feature of modern EFT approaches such as the
well-established and widely used infrared regularized \cite{Becher:1999he} and
extended-on-mass-shell (EOMS) \cite{Gegelia:1999gf,Fuchs:2003qc}
formulations of covariant baryon chiral perturbation theory (ChPT) as
well as the small-scale ($\epsilon$) expansion \cite{Hemmert:1997ye} within the covariant chiral EFT with explicit $\Delta
(1232)$ degrees of freedom, see, e.g., Ref.~\cite{Yao:2016vbz}.  
Consider, for example, the purely perturbative application of EOMS ChPT to pion-nucleon scattering.
\begin{figure}[tb]
	\includegraphics[width=0.5\textwidth]{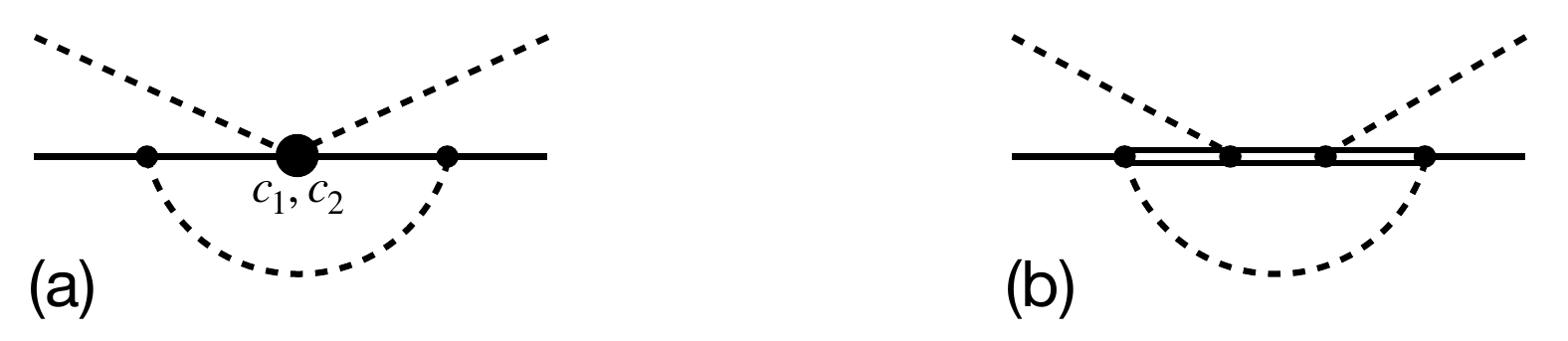}
	\caption{Examples of loop diagrams contributing to the $\pi N$ amplitude
          in covariant ChPT with a mismatch between the chiral order and the power of the ultraviolet divergence.
		Diagram (a) contributes at order $Q^4$ while diagram
                (b) with intermediate delta excitations shown by
                double lines appears at order $\epsilon^3$ in the small scale
                expansion. For remaining notation see Fig.~\ref{fig1}.}
	\label{Fig:Diagrams_piN}
\end{figure}
The one-loop diagram in Fig.~\ref{Fig:Diagrams_piN}~(a) with a single
insertion of vertices from the subleading
$\pi N$ Lagrangian $\mathcal{L}_{\pi N}^{(2)}$ is of order $\mathcal{O}(Q^4)$.
Nevertheless, to fully absorb the ultraviolet divergences and obtain a
renormalization-scale independent result, 
one has to add the counter terms of order $\mathcal{O}(Q^5)$.
Even more severe is the situation in the $\Delta$-full theory. For
example, the diagram shown in Fig.~\ref{Fig:Diagrams_piN}~(b) 
is of order $\epsilon^3$ in the $\epsilon$ counting with $m_\Delta - m
\sim \mathcal{O} (M_\pi)$, while its
ultraviolet divergent part has coefficients of orders up to
$\epsilon^9$. The residual scale dependence
does not invalidate the EFT expansion based on the power counting in the soft region:
Using dimensional regularization with the EOMS scheme, the remaining
logarithmic $\nu$-dependence of the considered amplitudes is a
higher-order effect.
Of course, convergence of the perturbative expansion will be spoiled
if one chooses the renormalization scale of the order of $\nu\sim m \,
e^{\Lambda_b/M_\pi}$.

Last but not least, we emphasize that it is, of course, possible to promote
certain contributions to lower orders in the Weinberg scheme in order to make the EFT expansion more efficient.
In particular, it might well be that the LECs $D_2$, whose value can,
in principle, be determined
in lattice-QCD simulations, will turn out to be larger than expected based on NDA.
It has also been argued that  in the $^1S_0$ channel, it might be preferable to promote the momentum
dependent $C_2$-term to LO \cite{Epelbaum:2015sha,Long:2013cya,Gasparyan:2023rtj}.
However, the reason for such promotions is a slow convergence of the
EFT expansion rather than RG-based arguments or formal inconsistencies of the scheme.

\subsection{Estimated magnitude of $D_2$}
\label{sec:SizeD2}

Since the numerical value of $D_2$  is unknown, we follow the approach of
Ref.~\cite{Cirigliano:2024ocg} by estimating it from the RG running, see
Eq.~(\ref{DressedOPEP}),
\beq
|D_2| \; \sim \; \frac{g_A^2 m^2}{6 \pi^2 F_\pi^2} C_0^2\; \equiv \;
\xi C_0^2\,,
\eeq
where $\xi \approx 0.27$. Various estimations of
$D_2$ can be summarized as follows:
\begin{itemize}
\item
  Cirigliano {\it et al.} \cite{Cirigliano:2024ocg}. \\
  The authors of this paper assert: {\it ''In Chiral EFT, for
    typical $\Lambda$ employed in calculations, $|\tilde C_0 | \sim
    1/m_\pi^2 \sim 5$~fm$^2$, and $\xi < 0.5$ predicts $|D_2 |
    \lesssim 10$~fm$^4$.''} They acknowledge that the range of
  $D_2$ they explore is comparable to that suggested in Ref.~\cite{Beane:2002xf},
  which uses the KSW power counting to describe NN scattering in the $^1$S$_0$
  channel. In the actual applications to nuclear matter, a smaller range of
  values with $|D_2 | \leq (5 F_\pi^4)^{-1} \approx 4.2$~fm$^4$ is employed.  
\item
  KSW choice of renormalization conditions.\\
  In the KSW scheme, the expected magnitude of the LO contact interaction
  can be read off from Eq.~(\ref{C0C2}) by setting $\mu \sim M_\pi$: $|C_0| \sim 4 \pi/(m M_\pi)
  \approx 3.8$~fm$^{2}$.  Using the actual numerical value of $\xi$,
  this leads to the estimation  $|D_2 | \sim 3.9$~fm$^4$, which
  is very close to the one adopted in Ref.~\cite{Cirigliano:2024ocg}. 
\item
  Weinberg's choice of renormalization conditions. \\
  For the choice of renormalization conditions outlined in
  sec.~\ref{sec:Weinberg}, one obtains from Eq.~(\ref{C0C2}) the
  estimation $|C_0| \sim 1.2$~fm$^2$, where the
  renormalization scale $\mu$ was set to a typical momentum cutoff
  $\mu \sim \Lambda = 450$~MeV used in chiral EFT
  calculations, see, e.g., \cite{Reinert:2017usi,Entem:2017gor,Ekstrom:2015rta}. This translates to the estimation
  $|D_2 | \sim 0.4$~fm$^4$.
\item
  Finite-cutoff chiral EFT as implemented in Ref.~\cite{Reinert:2017usi}. \\
  The semilocal momentum-space regularized (SMS) chiral NN potentials
  of Ref.~\cite{Reinert:2017usi} have been recently analyzed within a Bayesian
  statistical framework by the BUQEYE Collaboration and found to yield a regular convergence
  pattern for the cutoff values of $\Lambda = 450$ and $500$~MeV \cite{Millican:2025sdp}.  
  Taking the corresponding values for $\tilde C_{1S0}^{\rm np}$ from
  Table~2 of Ref.~\cite{Reinert:2017usi} and adjusting them to the
  convention used in this paper, we have $C_0 = (4 \pi)^{-1} \tilde
  C_{1S0}^{\rm np} = \{0.3, \,1.7\}$~fm$^2$ for $\Lambda = \{450, \,
  500\}$~MeV, which leads to the estimations $|D_2 | \sim
  \{ 0.03,\, 0.8\}$~fm$^4$.\footnote{Using the same considerations for
    the softest cutoff $\Lambda = 400$~MeV leads to $|D_2 | \sim
  0.25$~fm$^4$, while for the hardest cutoff $\Lambda = 550$~MeV a
  larger value of $|D_2 | \sim
  2.75$~fm$^4$ is obtained. We emphasize, however, that the SMS
  potentials with $\Lambda = 550$~MeV are highly nonperturbative and
  do not fully comply with the naturalness assumption for LECs
  \cite{Reinert:2017usi}. The strong $\Lambda$-dependence of $\tilde
  C_{1S0}^{\rm np}$ is probably caused by the non-perturbative
  treatment of subleading NN interactions within this framework.}  
  Alternatively, the size of $D_2$ can be estimated using the values
  of LECs accompanying momentum-dependent order-$Q^2$ contact interactions like
  $C_2$. Using the values quoted in Table 2 of
  Ref.~\cite{Reinert:2017usi} for $C_X$, where $X$ refers to a partial
  wave, the largest in magnitude value across all partial waves
  and cutoffs is $(4 \pi)^{-1} |C_X| = 1.0$~fm$^4$. 
\end{itemize}  
Based on the above considerations, we expect the estimation
\beq
\label{UsedValueD2}
|D_2 | \sim 1\text{ fm}^4
\eeq
to provide a realistic upper bound for this LEC in
calculations utilizing Weinberg's power counting.

\section{Chiral expansion of the two-pion-exchange-contact 3NF}
\label{sec:ChEFT3NF}

As already pointed out in the introduction, the $D_2 M_\pi^2$ two-nucleon operator
induces the corresponding interactions 
involving an even number of pions by virtue of chiral symmetry. Keeping only terms that are relevant
for this study, the effective heavy-baryon Lagrangian considered in 
Ref.~\cite{Cirigliano:2024ocg} has the form
\beq
\label{LagrCirigliano}
\mathcal{L} = \big( d_2^S \bar N N \bar NN + d_2^T  \bar N \vec \sigma N
\cdot \bar N\vec \sigma N \big) \langle \chi_+\rangle +
\big( f_2^S \bar N N \bar NN + f_2^T  \bar N \vec \sigma N
\cdot \bar N\vec \sigma N \big) \langle u\cdot u - (v \cdot u )^2\rangle\,,
\eeq
where $\langle \ldots \rangle$ denotes a trace in the flavor
space. Further, $u_\mu$ and $\chi_+$ are pion-field-dependent matrices in the flavor
space with the properties $\langle u\cdot u \rangle =
2 \partial_\mu\fet \pi \cdot \partial^\mu \fet \pi/F_\pi^2 +
\mathcal{O} (\fet \pi^4)$ and $\langle \chi_+ \rangle = 4 M_\pi^2 \big[1 -
\fet \pi^2/(2 F_\pi^2) + \mathcal{O} (\fet \pi^4) \big]$, see
Ref.~\cite{Bernard:1995dp} for details. In the rest frame of the nucleon, the
four-velocity vector $v_\mu$ takes the form $v_\mu = (1, \vec 0 \,
)$. Following Ref.~\cite{Cirigliano:2024ocg}, we express the LECs
$d_2^{S,T}$ and $f_2^{S,T}$ in terms of $D_2$ and $F_2$ via  
\beq
\label{d2andf2}
d_2^S = - d_2^T = - \frac{1}{32} D_2, \qquad
f_2^S = - f_2^T = - \frac{1}{32} F_2\,.
\eeq
Given that the RG equations for the LECs $F_2$ and $D_2$ are
similar \cite{Cirigliano:2024ocg}, we expect their magnitude to
be comparable and thus estimate $|F_2 |
\lesssim 1\text{ fm}^4$ in analogy with Eq.~(\ref{UsedValueD2}). 

The effective Lagrangian in Eq.~(\ref{LagrCirigliano}) gives rise to
the two-pion-exchange-contact 3NF at order $Q^6$ (i.e., N$^5$LO) shown in
Fig.~\ref{fig1}. Using dimensional regularization and employing
Eq.~(\ref{d2andf2}), the resulting 3NF has the form \cite{Cirigliano:2024ocg} 
\beq
\label{3NFNewClass}
V_{3N}^{(6)} = (1 - \vec \sigma_1 \cdot \vec \sigma_2) \big[ v^{(6)}_{D_2}
(q_3) +  v^{(6)}_{F_2} (q_3) \big] \; + \; \text{5 permutations}\,, 
\eeq
where the functions $v^{(6)}_{D_2}(q)$ and $v^{(6)}_{F_2}(q)$ are given by
\beqa
\label{vD2F2}
v^{(6)}_{D_2} (q) &=& \frac{3 g_A^2 D_2 }{256 \pi F_\pi^4} M_\pi^2 (2
M_\pi^2 + q^2) A(q) \; + \; \text{contact terms}\,, \nonumber \\
v^{(6)}_{F_2} (q) &=& \frac{3 g_A^2 F_2 }{512 \pi F_\pi^4}  (2
M_\pi^2 + q^2)^2 A(q) \; + \; \text{contact terms}\,,
\eeqa
and the loop function $A(q)$ is defined as
\beq
A(q) = \frac{1}{2q} \arctan \frac{q}{2M_\pi}\,.
\eeq
The scheme-dependent contact terms not shown explicitly in the first and second lines
of Eq.~(\ref{vD2F2}) are polynomials in $q^2$ of the zeroth and first
degrees, respectively, and can be absorbed into a redefinition
of the LECs $E$ \cite{Epelbaum:2002vt} and $E_i$ \cite{Girlanda:2011fh} that accompany the
purely short-range ($M_\pi$-dependent) N$^2$LO and
N$^4$LO 3NFs of type (f), see Fig.~\ref{fig1}.  

It is instructive to compare the strength of $V_{3N}^{(6)}$ in Eq.~(\ref{3NFNewClass}) to the
type-(e) 3NF contributions at lower orders N$^3$LO and
N$^4$LO shown in Fig.~\ref{fig1}, which, however, depend on the LO NN contact interactions, whose values
are scheme dependent. To have a more meaningful comparison, one can
alternatively examine pion-exchange diagrams of types (a), (b) or
(c), which do not involve scheme-dependent LECs and contain short-range
admixtures of types (d) and (e) much like the short-range part of the
$1\pi$-exchange potential in Eq.~(\ref{OPEP}). To be specific, consider
the two-pion-one-pion exchange diagrams of type (b), whose
contributions to the 3NF can be written in the form \cite{Krebs:2013kha}
\beqa
\label{two_pion_one_pion_general}
V_{3N}^{\rm 2\pi \mbox{-}1\pi} &=& 
 \frac{\vec \sigma_1 \cdot \vec q_1}{q_1^2 + M_\pi^2} \Big\{ \fet \tau_1
  \cdot \fet \tau_3  \; \left[ \vec \sigma_2 \cdot \vec q_3 \; \vec q_3 \cdot
    \vec q_1 \; F_1 (q_3)  + \vec \sigma_2 \cdot \vec q_3 \; F_2 (q_3) + 
  \vec \sigma_2 \cdot \vec q_1 \;  F_3 (q_3)  \right] + \fet \tau_1
  \cdot \fet \tau_2 \; [  \vec \sigma_3 \cdot \vec q_3 \;  \vec q_3 \cdot
    \vec q_1 \; F_4 (q_3) \nonumber \\
&& {} +  \vec \sigma_3 \cdot \vec q_1 \; F_5  (q_3) +
\vec \sigma_2 \cdot \vec q_3 \; \vec q_3 \cdot \vec q_1 \;  F_6 (q_3) +  
\vec \sigma_2 \cdot \vec q_3 \; F_7(q_3)
+  \vec \sigma_2 \cdot \vec q_1 \;  \vec q_3 \cdot \vec q_1 \; F_{8} (q_3)
+\vec \sigma_2 \cdot \vec q_1 \; F_{9}(q_3)]  \nonumber \\
&& {} + \fet \tau_1
  \times \fet \tau_2 \cdot \fet \tau_3 \left[ \vec \sigma_2 \times \vec \sigma_3
  \cdot \vec q_3 \;  (\vec q_3 \cdot
    \vec q_1 \; F_{10} (q_3)+ F_{11}(q_3))+\vec q_1 \times \vec q_3
    \cdot \vec \sigma_3 \; \vec q_3 \cdot \vec \sigma_2 \; F_{12}(q_3)\right]
  \Big\} \nonumber \\
&&{} + \; \text{5 permutations}
  \,. 
\eeqa
The N$^3$LO and N$^4$LO expressions for the functions $F_i (q)$, obtained 
using dimensional regularization, are given in Eqs.~(3.2) and (3.3) of
Ref.~\cite{Krebs:2013kha}. The spin-momentum operators in front of the 
functions $F_{3,5,9}$ resemble the form of the
$1\pi$-exchange. Expressing these operators in terms of $T_{1i}(\vec
q_1)$ via
\beq
\vec \sigma_1 \cdot \vec q_1 \, \vec \sigma_i \cdot \vec
q_1 = T_{1i}(\vec
q_1) + \frac{1}{3} (q_1^2 + M_\pi^2) \vec \sigma_1 \cdot \vec \sigma_i - \frac{1}{3}
M_\pi^2 \vec \sigma_1 \cdot \vec \sigma_i
\eeq
and cancelling $ q_1^2 +
M_\pi^2$ against the pion propagator in
Eq.~(\ref{two_pion_one_pion_general}), we read out the induced
two-pion-exchange-contact 3NFs in the form
\beqa
\label{3NFinduced}
V_{3N} & =&  \sum_{i=1}^3 V_i \; + \; \text{5 permutations}\nonumber \\
& =& \fet \tau_1 \cdot \fet \tau_3 \vec \sigma_1 \cdot \vec \sigma_2
\, v_1 (q_3) +
\fet \tau_1 \cdot \fet \tau_2 \vec \sigma_1 \cdot \vec \sigma_3
\, v_2 (q_3) +
\fet \tau_1 \cdot \fet \tau_2 \vec \sigma_1 \cdot \vec \sigma_2
\, v_3 (q_3) \; + \; \text{5 permutations}\,,
\eeqa
where the contributions to the functions $v_i (q)$
are given in terms of the corresponding $F_i(q)$ and read
\cite{Krebs:2013kha}\footnote{The total contribution of all diagrams
  of type (b) leads to a vanishing result for the function $F_9(q)$ at
  N$^3$LO \cite{Krebs:2013kha}. For the sake of comparison, we keep in $v_3^{(4)} (q)$ only
the contribution of the first N$^3$LO diagram of type (b) shown in
Fig.~\ref{fig1}, which can be found in Ref.~\cite{Bernard:2007sp}.}
\beqa
\label{vN3LO}
v_1^{(4)} (q) &=&  -\frac{g_A^4 }{768 \pi  F_\pi^6} \left[\left(8 g_A^2-4\right) M_\pi^2+\left(3 
    g_A^2-1\right) q^2\right] A(q) \,, \nonumber \\
v_2^{(4)} (q) &=&\frac{g_A^6 }{384 \pi
  F_\pi^6} q^2 A(q) \,, \nonumber \\
v_3^{(4)} (q) &=&\frac{g_A^4}{384 \pi
  F_\pi^6} (2 M_\pi^2 + q^2) A(q)\,, 
\eeqa
at N$^3$LO 
\beqa
\label{vN4LO}
v_1^{(5)} (q) &=&
-\frac{g_A^2 c_4 }{144 \pi^2 F_\pi^6 } \, \frac{L(q)}{4 M_\pi^2+q^2} \left[4 \left(4 g_A^2-1\right) 
M_\pi^4+\left(17 g_A^2-5\right) M_\pi^2 q^2+\left(4 g_A^2-1\right) 
q^4\right], \nonumber \\
v_2^{(5)} (q) &=&\frac{g_A^4 c_4 }{48
  \pi^2 F_\pi^6} q^2 L(q) \,, \nonumber \\
v_3^{(5)} (q) &=&\frac{g_A^4}{384 \pi^2 
F_\pi^6 } \, \frac{L(q) }{4 M_\pi^2+q^2} \Big[32 c_1 M_\pi^2 \left(3
  M_\pi^2+q^2
\right)-c_2 \left(16 M_\pi^4+16 M_\pi^2 q^2+3 q^4\right) \nonumber \\
&&{} -c_3 
\left(80 M_\pi^4+68 M_\pi^2 q^2+13 q^4\right)\Big],
\eeqa
at N$^4$LO. Here, we do not show terms polynomial in $q^2$, which can
be absorbed into shifts of the LECs $E$ at N$^3$LO and $E$, $E_i$ at N$^4$LO. Furthermore, $c_i$
denote the LECs of the subleading pion-nucleon Lagrangian, while the loop function
$L(q) $ is given by
\beq
L(q)  =  \frac{\sqrt{q^2 + 4 M_\pi^2}}{q} \log \frac{\sqrt{q^2 + 4
    M_\pi^2} + q}{2 M_\pi} \,.
\eeq
We emphasize that the decomposition of the 3NF into specific
topologies is ambiguous, see Ref.~\cite{Krebs:2012yv,Krebs:2013kha} for details. Moreover,
pion exchange potentials become scheme dependent starting from N$^3$LO, see a recent work
\cite{Springer:2025ojd} for a discussion. These ambiguities are,
however, of no relevance for the purpose of this paper. 

A close examination of the N$^3$LO,  N$^4$LO and N$^5$LO expressions
for the functions $v_i^{(4)} (q)$, $v_i^{(5)} (q)$ and $v^{(6)}_{D_2} (q)$,
$v^{(6)}_{F_2} (q)$ in Eqs.~(\ref{vN3LO}), (\ref{vN4LO}) and (\ref{vD2F2}),
respectively, reveals the well-known convergence issues with the
chiral expansion of pion-exchange potentials, which is plagued by
enhancements beyond NDA. In particular,
one-loop expressions involving the function $A(q)$ feature an
additional factor of $\pi$, while the N$^4$LO 
results for $v_1^{(5)} (q)$, $v_3^{(5)} (q)$ come with
unexpectedly large numerical coefficients, suggesting that the
frequently employed
estimation of the breakdown scale of ChPT from its upper bound \cite{Manohar:1983md}
$\Lambda_b \sim \Lambda_\chi = 4 \pi F_\pi$ is too
optimistic\footnote{Indeed, recent studies utilizing Bayesian statistics to infer the
breakdown scale of chiral EFT in the NN sector find $\Lambda_b
\sim 600-700$~MeV
\cite{Furnstahl:2015rha,Epelbaum:2019wvf,Millican:2025sdp}, see also
Ref.~\cite{Epelbaum:2014efa} for a related discussion.}.
For example, using our
estimation of $D_2$ in Eq.~(\ref{UsedValueD2}), one finds
\beq
\left|\frac{v_{D_2}^{(6)}(q)}{v_{3}^{(4)}(q)} \right|\; =\; \frac{9}{g_A^2} |D_2| F_\pi^2
M_\pi^2 \; \simeq \; 0.3
\eeq
instead of $\simeq 0.1$ as one would expect assuming the expansion
parameter $Q \sim 1/3$.  The situation is even more extreme for the N$^4$LO
contributions, which suffer from a double enhancement due to the
appearance of large numerical coefficients and large values of the LECs $c_{2,3,4}$ driven
by the $\Delta$-isobar \cite{Bernard:1996gq}, leading to, e.g.,
\beq
\label{StrongestEnhancement}
\left|\frac{v_{3}^{(5)}(0)}{v_{1}^{(4)}(0)} \right| \; =\; \frac{8 (6 c_1 -
  c_2 - 5 c_3) M_\pi}{(2 g_A^2 -1) \pi}\; \simeq \; 2.6\,,
\eeq
where we use the values of $c_1 = -1.10$~GeV$^{-1}$,  
$c_2 = 3.57$~GeV$^{-1}$ and $c_3 = -5.54$~GeV$^{-1}$ from the
N$^3$LO$^{NN}$ determination of Ref.~\cite{Hoferichter:2015tha}.  
We emphasize that this extreme example should not be
misinterpreted as a general statement regarding convergence of the chiral
expansion for long-range nuclear potentials. For example, the results
for the two-pion exchange 3NF of type (a) show a reasonable convergence
pattern consistent with the expansion parameter of $Q \sim
1/3$, see Fig.~4
of Ref.~\cite{Endo:2024cbz}. The convergence of the chiral expansion
for long-range 3N potentials in coordinate space, generated by the pion-exchange topologies
(a), (b) and (c), is discussed in detail in
Refs.~\cite{Krebs:2012yv,Krebs:2013kha,Epelbaum:2014sea,Krebs:2018jkc},
showing that the subleading loop corrections at N$^4$LO  are, in general, 
expected to yield sizable contributions. 

\section{$t$-channel pion loops: Lessons from the NN sector}
\label{sec:ChEFT2NF}

The above considerations provide a strong motivation to examine the
situation in the two-nucleon sector. This motivation is twofold:
First, the expressions for the 3NF obtained by Cirigliano {\it et al.} \cite{Cirigliano:2024ocg}
closely resemble the subleading contributions to the two-pion exchange
NN potential as follows from the similarity of the Lagrangian in
Eq.~(\ref{LagrCirigliano}) and
\beq
\mathcal{L}_{\pi N}^{(2)} = \bar N \left( c_1 \langle \chi_+ \rangle +
  c_2 (v \cdot u)^2 + c_3 u \cdot u + \ldots \right) N\,.
\eeq
Secondly, in contrast to the 3NF, which so far has only been
applied at the N$^2$LO accuracy level, NN scattering has been extensively
studied at high expansion orders, which allows one to have  
a more complete and reliable assessment of the convergence rate of chiral
EFT and of the efficiency of specific renormalization schemes.  

In the static limit with $m \to \infty$, the NN potential from multi-pion exchange
is strictly local and can be expressed in terms of six
scalar functions $v_{C,S,T} (q)$ and $w_{C,S,T} (q)$ via 
\beqa
V (\vec q \, ) &=& v_C (q) + \vec \sigma_1 \cdot \vec \sigma_2 \, v_S (q) +  \vec
\sigma_1 \cdot \vec q \, \vec \sigma_2 \cdot \vec q \, v_T (q)
+ \fet \tau_1 \cdot \fet \tau_2 \left[ w_C (q) + \vec \sigma_1 \cdot
  \vec \sigma_2 \, w_S (q) +  \vec
\sigma_1 \cdot \vec q \, \vec \sigma_2 \cdot \vec q \, w_T (q) \right] . 
\eeqa
The LO (i.e., order-$Q^0$) term is given by the $1 \pi$-exchange in Eq.~(\ref{1pi}),
while the $2 \pi$-exchange contributions at NLO ($Q^2$) and N$^2$LO
($Q^3$), calculated using dimensional regularization, have the form \cite{Kaiser:1997mw,Epelbaum:1998ka}
\beqa
\label{TPEPNN}
w_C^{(2)} (q) &=&
- \frac{ 1}{384 \pi^2 F_\pi^4}\,
L(q) \, \left[4M_\pi^2 (5g_A^4 - 4g_A^2 -1)
+ q^2(23g_A^4 - 10g_A^2 -1)
+ \frac{48 g_A^4 M_\pi^4}{4 M_\pi^2 + q^2} \right], \nonumber \\
v_S^{(2)} (q) &=&- q^2 v^{(2)}_T(q) \; = \; \frac{3 g_A^4}{64 \pi^2
  F_\pi^4} \,q^2 L(q) \,, \nonumber \\
v_C^{(3)} (q) &=&-\frac{3g_A^2}{16\pi F_\pi^4}  \left[2M_\pi^2(2c_1
  -c_3) -c_3 q^2 \right] (2M_\pi^2+q^2) A(q) \,,
\nonumber \\
w_S^{(3)} (q) &=&- q^2 w^{(3)}_T(q) \; = \;
\frac{g_A^2 c_4}{32\pi F_\pi^4} \, q^2 (4M_\pi^2 + q^2) A(q)\,,
\eeqa
where only terms non-polynomial in $q^2$ are shown. Higher-order
corrections to the two- and three-pion exchange NN potentials up to
N$^4$LO and, in part, even N$^5$LO can be found in
Refs.~\cite{Kaiser:1999ff,Kaiser:2001pc,Kaiser:2001at,Entem:2014msa,Kaiser:1999jg,Kaiser:2001dm,Springer:2025ojd},
but they are of no relevance for our discussion. It is worth noting
that the expression for $v_C^{(3)} (q)$ coincides  (upon renaming the LECs $c_{1,3}$) with the
functions $v_{D_2}^{(6)}$   $v_{F_2}^{(6)}$ in Eq.~(\ref{vD2F2}) derived by
Cirigliano {\it et al.}~\cite{Cirigliano:2024ocg}, as it
was already mentioned at the beginning of this section. 

\begin{figure}[tb]
  \includegraphics[width=0.8\textwidth]{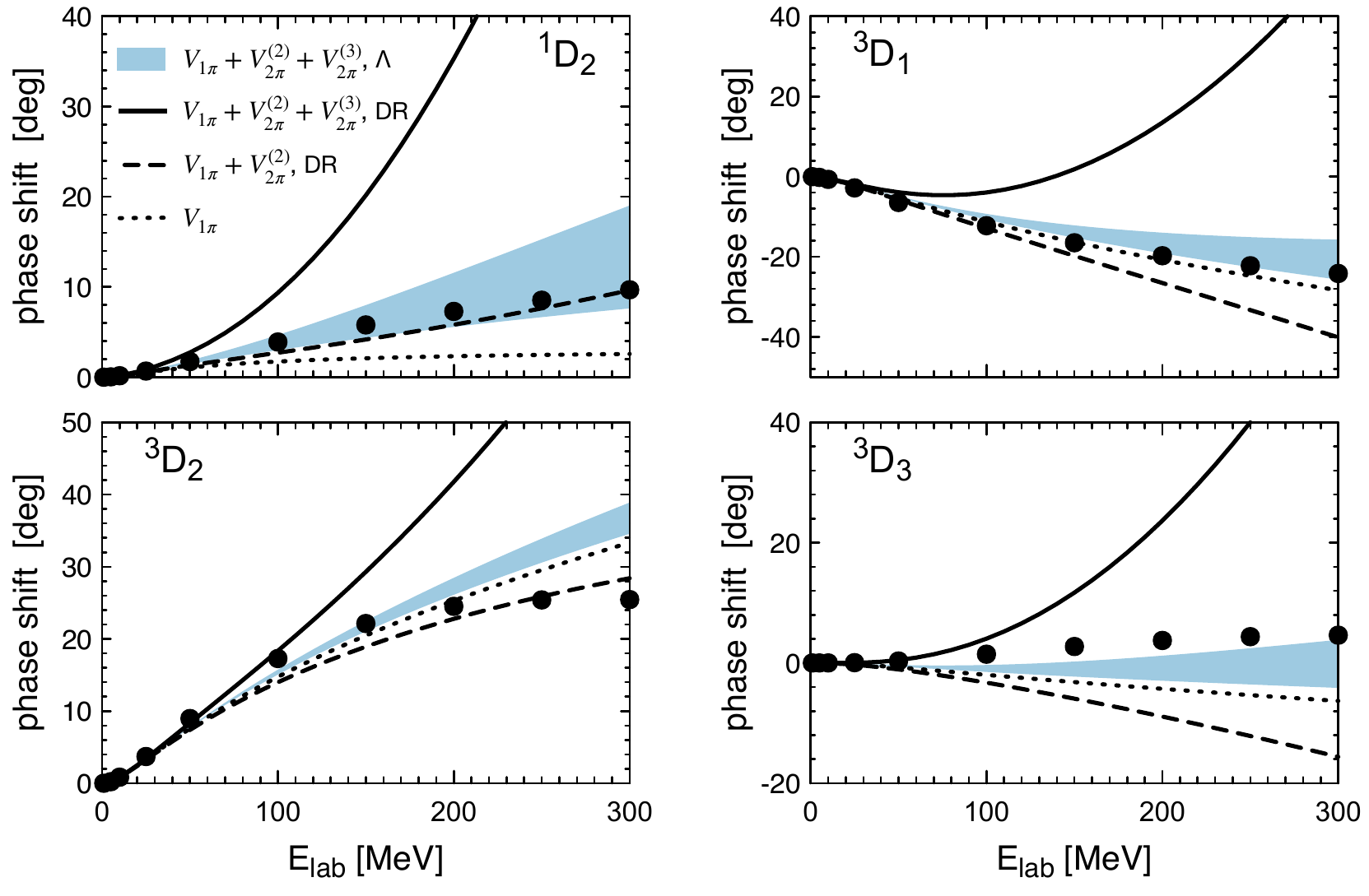}
  \vspace{-0.35cm}
	\caption{ChPT predictions for neutron-proton D-wave
          phase shifts. Dotted lines are LO results based on
          the $1\pi$-exchange potential, while dashed and solid lines
          emerge from taking into account the NLO and N$^2$LO
          $2\pi$-exchange in Eq.~(\ref{TPEPNN}) calculated using dimensional
          regularization. Blue bands show the N$^2$LO predictions
          after removing the short-range components of the
          $2\pi$-exchange by imposing a sharp cutoff $\Lambda =
          500-800$~MeV in the spectral integral. Filled circles are
          empirical phase shifts of the Nijmegen partial wave analysis
        \cite{Stoks:1993tb}. Figure adapted from Refs.~\cite{Epelbaum:2024gfg,Epelbaum:2003gr}.}
              \label{Fig:DWaves}
\end{figure}

As one can see from Eq.~(\ref{TPEPNN}), the chiral expansion of the $2\pi$-exchange NN potential suffers from
the same kind of enhancements as discussed in
sec.~\ref{sec:ChEFT3NF} for the type-(e) 3NF. The situation is, in fact, even worse for
$v_C^{(3)} (q)$, which shows a triple enhancement due to an additional
factor of $\pi$ in the numerator, the appearance of large numerical coefficients and
the large magnitude of the LEC $c_3$. This has important
phenomenological implications, which can be most cleanly observed in
peripheral NN scattering. For D- and higher waves, the centrifugal
barrier filters out the contributions of the LO and NLO contact
interactions, and the corresponding phase shifts and mixing angles are
determined solely by the $1\pi$- and $2\pi$-potentials given in
Eqs.~(\ref{1pi}) and (\ref{TPEPNN}). Moreover, for not too large cutoff values,
the amplitude can be well represented by the Born
approximation. Accordingly, NN scattering in D-waves can be used
to directly probe the ChPT predictions for the $2\pi$-exchange in
Eq.~(\ref{TPEPNN}). These ideas were taken up in Ref.~\cite{Kaiser:1997mw}, leading
to the results visualized in
Fig.~\ref{Fig:DWaves}.\footnote{ChPT predictions shown in
  Fig.~\ref{Fig:DWaves} do not include relativistic corrections and
  reducible $2\pi$-exchange contributions from the iterated $1\pi$-exchange taken into account in
Ref.~\cite{Kaiser:1997mw}, which appear to be numerically suppressed.}  
For $E_{\rm lab} \gtrsim 100$~MeV, all D-wave phase shifts are overestimated
at N$^2$LO, signalling an unphysically strong attraction generated by $v_C^{(3)}
(q)$, which, albeit less pronounced, is also visible in F-waves
\cite{Kaiser:1997mw}. The failure of ChPT to describe peripheral NN
scattering can be traced back to enhanced scheme-dependent short-range components of the
$2\pi$-exchange captured by the dimensionally regularized $t$-channel
loop integrals in Eq.~(\ref{TPEPNN}) \cite{Epelbaum:2003gr}. To see
this it is instructive to express the functions $v_i(q)$ and $w_i(q)$,
with $i = \{C,S,T\}$, using the (twice subtracted) dispersion relations
via
\beqa
\label{Dispersive}
v_i (q) = \bigg[\frac{2 q^4}{\pi} \int_{2 M_\pi}^{\Lambda} 
\frac{d \mu}{\mu^3} \frac{\rho_i (\mu)}{\mu^2 +
  q^2}
+
\int_{\Lambda}^\infty 
\frac{d \mu}{\mu^3} \frac{\rho_i (\mu)}{\mu^2 + q^2}\bigg]
- P_i(q^2) \quad  &\Rightarrow& \quad 
v_{i}^\Lambda  (q) := \frac{2 q^4}{\pi} \int_{2 M_\pi}^{\Lambda} 
\frac{d \mu}{\mu^3} \frac{\rho_i (\mu)}{\mu^2 +
  q^2}, \phantom{XX}\\
w_i (q) = \bigg[\frac{2 q^4}{\pi} \int_{2 M_\pi}^{\Lambda} 
\frac{d \mu}{\mu^3} \frac{\eta_i (\mu)}{\mu^2 +
  q^2}
+
\int_{\Lambda}^\infty 
\frac{d \mu}{\mu^3} \frac{\eta_i (\mu)}{\mu^2 + q^2}\bigg]
- P_i '(q^2)  \quad &\Rightarrow& \quad 
w_{i}^\Lambda  (q) := \frac{2 q^4}{\pi} \int_{2 M_\pi}^{\Lambda} 
\frac{d \mu}{\mu^3} \frac{\eta_i (\mu)}{\mu^2 +
  q^2} ,\nonumber
\eeqa
where $\rho_i (\mu) = {\rm Im} \left[ v_i (0^+ - i \mu ) \right]$,  $\eta_i (\mu) = {\rm Im} \left[ w_i (0^+ - i \mu ) \right]$
and $P_i(q^2)$ and $P_i'(q^2)$ are first-degree polynomials, which 
can be absorbed into a redefinition of the NN contact interactions at
LO and NLO and do not contribute to D- and higher partial waves. The functions
$v_i^\Lambda (q)$ and $w_i^\Lambda (q)$ can be chosen as an alternative
definition of the $2\pi$-exchange potential. On the conceptual level,
both definitions are equally valid in chiral EFT if $\Lambda \gtrsim
\Lambda_b$ and correspond to specific choices of renormalization
conditions, which differ from each other by finite pieces of (an
infinite set of) contact interactions. For rapidly convergent
expansion series showing no enhancements beyond NDA, both
choices may be expected to yield comparable results when calculating 
observables. However, as mentioned above,  $v_C^{(3)} (q)$ appears to
be strongly enhanced with, e.g., $| v_C^{(3)} (M_\pi) / w_C^{(2)}
(M_\pi) | \simeq 4$. Accordingly, using dimensional regularization to calculate the
$2\pi$-exchange potential at N$^2$LO results in enhanced NN
contact interactions, stemming from the second term in the square
brackets in Eq.~(\ref{Dispersive}). Contrary to the enhanced
long-range $2\pi$-exchange interaction generated by the low-$\mu$ part
of the spectrum, the enhanced short-range terms induced by using
dimensional regularization 
do not represent a meaningful prediction of chiral EFT, since the
chiral expansion for the spectral functions $\rho_i (\mu)$ is not expected to converge for $\mu
\gtrsim \Lambda_b$. Under these circumstances, it is preferable to
employ $v_i^{\Lambda} (q)$ and $w_i^{\Lambda} (q)$  with $\Lambda \sim
\Lambda_b$ instead of 
$v_i (q)$ and $w_i (q)$ in order to avoid an artificial enhancement of
higher-order NN contact interactions. The light-shaded blue bands in
Fig.~\ref{Fig:DWaves} show the N$^2$LO predictions for D-wave phase
shifts based on $v_i^{\Lambda} (q)$ and $w_i^{\Lambda} (q)$ with
$\Lambda = 500 - 800$~MeV, which show a significant improvement compared to
the NLO results and agree well with empirical values\footnote{Apart
  from the strongly enhanced subleading 
isoscalar-scalar potential, the chiral expansion of the long-range
part of the $2\pi$-exchange shows a reasonable convergence pattern up
through N$^4$LO, see Figs.~6 and 7 of Ref.~\cite{Reinert:2017usi}.}. In contrast,
dimensional regularization leads in this case to an
inefficient choice of renormalization conditions, which requires
a promotion of $\mathcal{O}(Q^4)$ contact terms to N$^2$LO in order to
compensate for artificially enhanced short-range interactions at order
$Q^4$ and beyond  and
thus does not allow one to fully benefit from the predictive power of
chiral EFT \cite{Epelbaum:2024gfg}.

\section{Nuclear matter estimations}
\label{sec:EoS}

After these preparations, we are now in the position to estimate the
impact of the new class of 3NF in Eqs.~(\ref{3NFNewClass}) and
(\ref{vD2F2}) on the EoS for neutron and symmetric nuclear
matter. To this aim, we follow the approach of
Ref.~\cite{Cirigliano:2024ocg} and calculate the
energy per particle contributions to the pure neutron matter, $E_{\rm
  NM}$, and symmetric nuclear 
matter, $E_{\rm SM}$, at the Hartree-Fock level as detailed in
Refs.~\cite{Tews:2012fj,Cirigliano:2024ocg}, i.e., by considering 
nucleons in the Fermi sea interacting via 3NFs introduced in sec.~\ref{sec:ChEFT3NF}.

We first use the same scheme and the same numerical values for 
$D_2$ and $F_2$ as done in Ref.~\cite{Cirigliano:2024ocg}. 
\begin{figure}[tb]
	\includegraphics[width=0.95\textwidth]{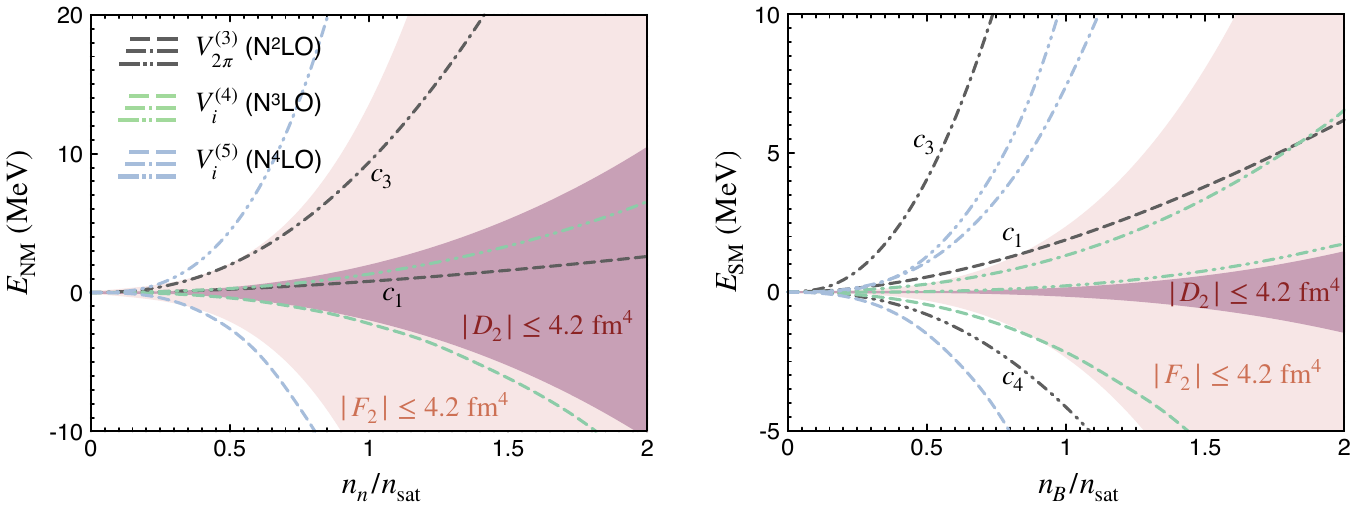}
	\caption{Energy per particle versus density in units of the
          saturation density $n_{\rm sat}$ from 
          dimensionally regularized 3NFs in neutron matter (left) and symmetric nuclear
          matter (right), calculated at the
          Hartree-Fock level. Black dashed, dashed-dotted and
          dashed-double-dotted lines show the effects from the $c_1$-, $c_3$- and
          $c_4$-contributions to the N$^2$LO 3NF of type (a),
          respectively. Green (blue) dashed, dashed-dotted and
          dashed-double-dotted lines refer to the $v_1$-,  $v_2$- and
          $v_3$-contributions to the 3NF of
          type (e) at N$^3$LO (N$^4$LO) specified in
          Eq.~(\ref{vN3LO}) (in Eq.~(\ref{vN4LO})). Light-shaded and dark-shaded bands show an
          estimation of the N$^5$LO corrections of type (e) driven by the LECs $F_2$ and $D_2$, respectively.
          These bands are obtained using the expressions for
          $v_{F_2}^{(6)}(q)$ and  $v_{D_2}^{(6)}(q)$ in
          Eq.~(\ref{vD2F2}), including the corresponding polynomial terms, and
          employing the range of $|D_2|, \, |F_2|
          \leq 4.2$~fm$^4$ advocated by Cirigliano {\it et al.}
          \cite{Cirigliano:2024ocg}.}
	\label{Fig:MatterCirigliano}
\end{figure}
In  Fig.~\ref{Fig:MatterCirigliano}, we show the
contributions to the neutron and symmetric nuclear matter EoS from
the unregularized N$^2$LO 3NF of type (a)
\cite{Epelbaum:2002vt},
\beqa
\label{TreeLevel}
V_{3N}^{2 \pi} &=& \frac{g_A^2}{8 F_\pi^4} \frac{\vec \sigma_1 \cdot \vec
q_1\, \vec \sigma_3 \cdot \vec
q_3}{(q_1^2 + M_\pi^2)(q_3^2+M_\pi^2)}
\Big[\fet \tau_1 \cdot \fet \tau_3 \big( - 4 c_1 M_\pi^2 + 2 c_3 \vec
q_1 \cdot \vec q_3 \big) + c_4 \fet \tau_1 \times \fet \tau_3 \cdot
\fet \tau_2 \vec q_1 \times \vec q_3 \cdot \vec \sigma_2
\Big]\nonumber\\
&+& \text{5 permutations}\,,
\eeqa
and from the considered N$^{3\text{-}5}$LO 3NFs of type (e) specified in
Eqs.~(\ref{3NFNewClass}), (\ref{vD2F2}), (\ref{3NFinduced}),
(\ref{vN3LO}) and (\ref{vN4LO}). For the LECs $c_i$, we employ
the values  $c_1 = -1.10$~GeV$^{-1}$,  
$c_2 = 3.57$~GeV$^{-1}$, $c_3 = -5.54$~GeV$^{-1}$ and $c_4 = 4.17$~GeV$^{-1}$ \cite{Hoferichter:2015tha}. 
Notice that for the N$^5$LO contributions, we also include
the polynomial in $q_3^2$ terms not shown explicitly in
Eq.~(\ref{vD2F2}) and, as already mentioned, employ the same range of the
LECs $D_2$ and $F_2$ as done in Ref.~\cite{Cirigliano:2024ocg}. The  
results shown by the black lines
and bands  in Fig.~(\ref{Fig:MatterCirigliano})  agree with the ones given in that paper. In particular,
we find very large N$^5$LO contributions driven by the
momentum-dependent $F_2$-vertex, the analogon of the $c_3$-interaction
in the single-nucleon sector. However, even larger effects are observed from 
the N$^4$LO 3NFs caused by the numerically enhanced functions
$v_i^{(5)}(q)$ in Eq.~(\ref{vN4LO}). 

The results shown in Fig.~(\ref{Fig:MatterCirigliano}) do, in
our view, not properly reflect the convergence pattern of
chiral EFT for neutron/nuclear matter in the Weinberg
scheme. 
To obtain more realistic estimates, we  
employ a local regulator for pion-exchange
interactions along the lines of Refs.~\cite{Reinert:2017usi,Epelbaum:2019zqc}. Specifically, the tree-level expressions in
Eq.~(\ref{TreeLevel}) are regularized via a replacement
\beq
\label{RegOPEP}
\frac{1}{(q_1^2 + M_\pi^2)} \, \frac{1}{(q_3^2 + M_\pi^2)}  \quad
\longrightarrow \quad
\frac{e^{-\frac{q_1^2+M_\pi^2}{\Lambda^2}}}{(q_1^2 + M_\pi^2)} \, \frac{e^{-\frac{q_3^2+M_\pi^2}{\Lambda^2}}}{(q_3^2 + M_\pi^2)} \,.
\eeq
For the $t$-channel pion loop integrals in the type-(e) contributions,
we use the once- and twice-subtracted dispersion relations with a local
Gaussian regulator as done in the SMS NN potentials of Ref.~\cite{Reinert:2017usi}, i.e., we
replace the functions $v_i^{(4)}(q)$, $v_i^{(5)}(q)$ and $v_{D_2, 
  F_2}^{(6)}(q)$ with\footnote{We do not employ here
  additional subtractions to ensure that the corresponding $r$-space
  potentials vanish at the origin as done in
  Ref.~\cite{Reinert:2017usi}. Such subtraction terms diverge in the
  limit $\Lambda \to \infty$, which complicates the comparison with
  the results of Ref.~\cite{Cirigliano:2024ocg}. These additional
  subtractions are, of course, absorbable into redefinition of the LECs $E$
  and $E_i$.}
\beqa
\label{RegTPEP}
v_{i}^{\Lambda  (4)} (q) &=& - \frac{2 q^2}{\pi} \int_{2 M_\pi}^\infty
\frac{d\mu}{\mu} \, \frac{e^{-\frac{q^2+M_\pi^2}{2\Lambda^2}}}{\mu^2 +
  q^2} \, {\rm Im} \Big[v_i^{(4)}\big(0^+ - i \mu\big)\Big], \nonumber \\ 
v_{i}^{\Lambda (5)} (q) &=&  \frac{2 q^4}{\pi} \int_{2 M_\pi}^\infty
\frac{d\mu}{\mu^3} \, \frac{e^{-\frac{q^2+M_\pi^2}{2\Lambda^2}}}{\mu^2 +
  q^2} \, {\rm Im} \Big[v_i^{(5)}\big(0^+ - i \mu\big)\Big], \nonumber
\\
v_{D_2, 
  F_2}^{\Lambda (6)} (q) &=&  \frac{2 q^4}{\pi} \int_{2 M_\pi}^\infty
\frac{d\mu}{\mu^3} \, \frac{e^{-\frac{q^2+M_\pi^2}{2\Lambda^2}}}{\mu^2 +
  q^2} \, {\rm Im} \Big[v_{D_2, 
  F_2}^{(6)}\big(0^+ - i \mu\big)\Big].
\eeqa
Notice that the regulators in Eqs.~(\ref{RegOPEP}) and
(\ref{RegTPEP}) are chosen to preserve the analytic structure
of the corresponding potentials and do not affect the pion pole and
the $t$-channel cut generated by the $2\pi$-exchange.  
When taking the $\Lambda \to \infty$ limit, the dispersive representation
in Eq.~(\ref{RegTPEP}) reduces to the expressions given in
Eqs.~(\ref{vD2F2}),  (\ref{vN3LO}) and (\ref{vN4LO}) up to 
polynomials in $q^2$. 
These polynomial terms lead to finite shifts of the LEC $E$ for the N$^3$LO
expressions and of the LECs $E$, $E_i$ in the case of the 
N$^4$LO and N$^5$LO 3NFs.

\begin{figure}[tb]
	\includegraphics[width=0.95\textwidth]{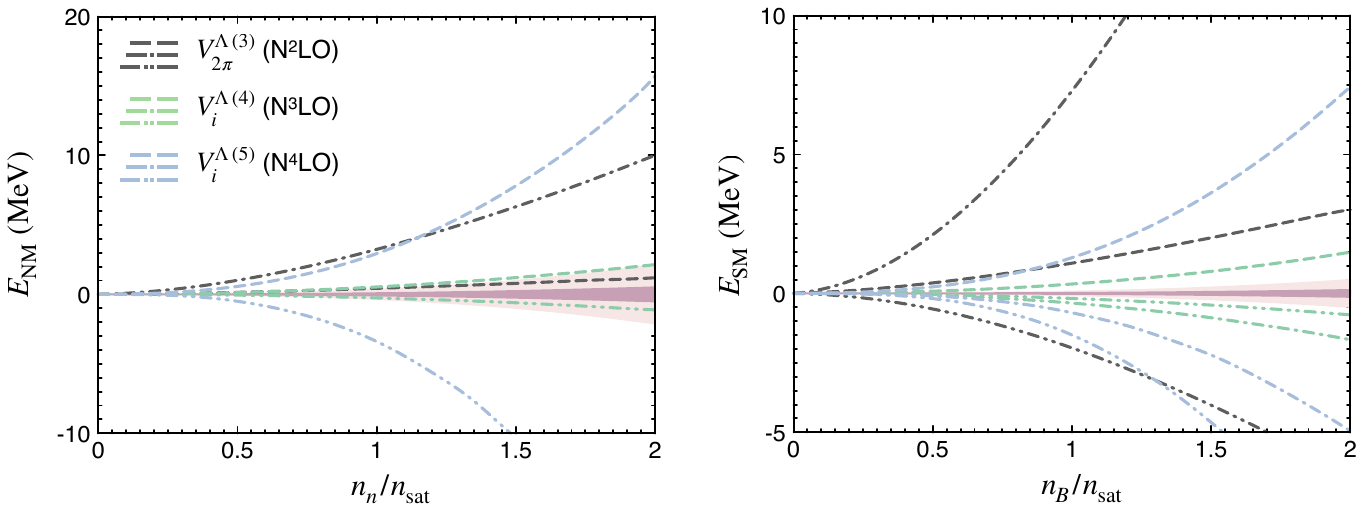}
	\caption{Same as in Fig.~\ref{Fig:MatterCirigliano} but using regularized
          3NF expressions as defined in Eqs.~(\ref{RegOPEP}) and
          (\ref{RegTPEP}) with $\Lambda = 500$~MeV. For the LECs $D_2$ and $F_2$, we consider the range
        of $|D_2|, \, |F_2|  \leq 1$~fm$^4$ as discussed in sec.~\ref{sec:SizeD2}.}
	\label{Fig:MatterOur}
      \end{figure}
      
In Fig.~\ref{Fig:MatterOur}, we show our results for pure neutron and
symmetric nuclear matter using regularized expressions for the 3NF
with $\Lambda = 500$~MeV and 
employing the range $|D_2|, |F_2| \leq 1$~fm$^4$ estimated  in
sec.~\ref{sec:SizeD2}. The usage of a smooth Gaussian-type regulator
has an obvious kinematical effect of suppressing the contributions to the energy per particle at
large densities when the Fermi momentum becomes comparable to the
cutoff value. This effect becomes clearly visible by comparing the tree-level
contributions shown by black lines in Figs.~\ref{Fig:MatterCirigliano} and \ref{Fig:MatterOur}. 
For loop contributions, the employed regularization also 
removes scheme-dependent short-range interactions induced by the
application of dimensional regularization, which cannot be reliably predicted
in chiral EFT and thus have to be regarded as artifacts of that
regularization scheme. As argued in sec.~\ref{sec:ChEFT2NF}, this issue becomes
particularly important for slowly convergent series. The much stronger
reduction of contributions from 3NFs involving pion loops compared to
those from tree-level 3NFs caused by the regulator indeed suggests that the
irregular convergence of the chiral expansion in
Fig.~(\ref{Fig:MatterCirigliano}) is largely caused by such 
short-distance artifacts. In fact, the difference between the results
shown in Figs.~\ref{Fig:MatterCirigliano} and \ref{Fig:MatterOur}
closely resembles the difference between the solid lines and light-shaded
bands in Fig.~\ref{Fig:DWaves}. We further emphasize that the
effects of the N$^{4\text{-}6}$LO 3NFs relative to those of the
dominant N$^2$LO 3NF in Fig.~\ref{Fig:MatterOur} are still
overestimated at high densities, since we do not employ any regulator
on the short-range part of the 3NFs of type (e).   

Our estimations for the new class of N$^5$LO 3NFs show a drastic reduction in
magnitude compared to those of Ref.~\cite{Cirigliano:2024ocg}. For
example, at the saturation density, our estimates of the contributions to the energy per
particle generated by the $D_2$ ($F_2$) 3NF appear to be $\sim 8$
($33$) times smaller for symmetric nuclear matter and 
$\sim 22$ ($42$) times smaller for pure neutron matter. This drastic reduction is
mainly caused by using the smaller (and in our view more adequate)
range of values for the unknown LECs $D_2$ and $F_2$ as
discussed in sec.~\ref{sec:SizeD2} and by removing spurious short-range
components from dimensionally regularized pion loop integrals in the $t$-channel.   
Our results show that using the finite-cutoff chiral EFT formulation
of sec.~\ref{sec:FiniteCutoff}, the ``new class'' of N$^5$LO 3NFs identified by Cirigliano
{\it et al.} \cite{Cirigliano:2024ocg} is expected to yield visible/small
contributions to the energy per particle of neutron/symmetric nuclear
matter, which for all considered densities are much smaller in magnitude than the dominant
contributions stemming from the N$^2$LO 3NF $\propto c_{3,4}$.    

Finally, we emphasize that all estimations considered here should be
understood to be of 
qualitative nature only as they are limited to a small subset of 3NF
contributions and make use of the simplistic Hartree-Fock
approximation with the Fermi gas reference state. Still, we believe
that the resulting pattern with rather sizable 3NF contributions generated
at N$^4$LO is robust, and it appears to be qualitatively consistent with
the conclusions of Refs.~\cite{Krebs:2012yv,Krebs:2013kha,Epelbaum:2014sea,Krebs:2018jkc}.

\section{Summary and conclusions}
\label{sec:Summary}

In this paper, we have scrutinized the arguments used in
Ref.~\cite{Cirigliano:2024ocg} to justify the need to promote 3NF diagrams
involving quark-mass/momentum dependent $\pi\pi$NN vertices from N$^5$LO to
N$^3$LO and reevaluated the impact of this new class of 3NFs on
the equation of state of neutron and symmetric nuclear  matter. 
Our main conclusions are summarized below.
\begin{itemize}
\item
As explained in detail in sec.~\ref{sec:ScalingD2}, RG arguments do
{\it not} justify the need to promote the quark-mass-dependent NN
contact interaction $\propto D_2$ to LO (unless the corresponding
logarithmic scale is artificially increased to nonsensically large
values to generate large logarithms\footnote{Such choice of
  renormalization conditions would also destroy convergence of
 e.g., the purely perturbative covariant formulations of baryon ChPT in the
  single-nucleon sector
  \cite{Becher:1999he,Gegelia:1999gf,Fuchs:2003qc} and of perturbative
  series in renormalizable theories.}).  
\item
  The scaling and magnitude of $D_2$ depend on the scheme/choice
  of renormalization conditions.
  The LEC $D_2$ is expected to scale according to NDA with $D_2 M_\pi^2 \sim
\mathcal{O} (Q^2)$
  in
Weinberg's power counting scheme. Using RG arguments, we estimate
$D_2$ to be of the order of $|D_2| \lesssim 1$~fm$^4$ when
using the finite-cutoff formulation of chiral EFT outlined in sec.~\ref{sec:FiniteCutoff}. The enhancement of the
operator $D_2 M_\pi^2$ by two inverse powers of the expansion
parameter relative to NDA assumed
by Cirigliano {\it  et al.} \cite{Cirigliano:2024ocg}, $D_2 M_\pi^2
\sim \mathcal{O} (1)$,
corresponds to the KSW choice of renormalization conditions. The
range of values of $D_2$ and $F_2$ considered in that paper, $|D_2|, |F_2| \leq
4.2$~fm$^4$, also closely resembles the values used in calculations
based on the KSW scheme \cite{Beane:2002xf}. The KSW approach is,
however, known to suffer from convergence issues already in the
two-nucleon sector \cite{Cohen:1998jr,Fleming:1999ee}.
We also do not expect the employed Hartree-Fock approximation to
provide adequate results for the EoS of neutron/nuclear matter within the KSW scheme.
\item
To gain insights into the convergence pattern of chiral EFT for
the type-(e) 3NF and into the relative importance of the N$^5$LO interactions
derived by Cirigliano {\it et al.}  \cite{Cirigliano:2024ocg}, we have isolated the parameter-free contributions of
the same type at orders N$^3$LO and N$^4$LO that are induced by the
two-pion-one-pion-exchange diagrams of type (b). The corresponding
expressions for the two-pion exchange show, as expected, the same type
of enhancement as observed in the two-nucleon sector, which is
caused by the appearance of large numerical factors and large
values of the subleading pion-nucleon LECs $c_{3,4}$. Under such
circumstances, the use of the renormalization scheme based on
dimensional regularization with the MS or $\overline{\text{MS}}$ scheme to compute
loop integrals appearing in two-pion exchange diagrams is well known to be inefficient \cite{Kaiser:1997mw,Epelbaum:2003gr}. A more efficient
and commonly used scheme relies on a cutoff regularization of the
corresponding spectral integrals, which allows one to avoid artificial
enhancement of short-range interactions \cite{Epelbaum:2003gr,Epelbaum:2014efa,Epelbaum:2024gfg}. 
This scheme is used, e.g., in the chiral EFT potentials of Refs.~\cite{Epelbaum:2004fk,Entem:2003ft,Gezerlis:2014zia,Ekstrom:2013kea,Ekstrom:2015rta,Entem:2017gor,Reinert:2017usi,Reinert:2020mcu}. 
\item
Inspired by the state-of-the-art SMS chiral NN potentials of Refs.~\cite{Reinert:2017usi,Reinert:2020mcu}, we employed a
Gaussian-type regulator in the dispersive representation of the 
type-(e) 3NFs with $\Lambda = 500$~MeV. Using the range of values
$|D_2|, \, |F_2|  \leq 1$~fm$^4$, the estimated contributions of the
new class of 3NFs to the energy per particle in neutron and symmetric
nuclear matter are found to be drastically reduced compared to the
estimations of Ref.~\cite{Cirigliano:2024ocg}, with our results at the
saturation density being 
up to $40$ times smaller. The estimated $D_2$ and $F_2$
contributions to the EoS of neutron/nuclear matter are found to
be much smaller than the ones generated by the leading 3NF $\propto
c_{3,4}$. 
\item
While the N$^5$LO 3NFs considered in Ref.~\cite{Cirigliano:2024ocg} are
estimated to have small effects in neutron/nuclear matter, we find
rather large type-(e) 3NF contributions at the subleading order (N$^4$LO)\footnote{Notice that we have only considered a small
  fraction of diagrams contributing to the type-(e) topology, so that
  the statements about convergence can only be considered as
  indicative.}. This is not surprising given the close similarity of
these 3NFs with the subleading two-pion exchange NN potential,
which is known to be strongly enhanced by the appearance of large
numerical coefficients. A similar type of
enhancement occurs for the considered 3NF diagrams of type-(e).  
On the other hand, the very strong attractive isoscalar-scalar
$2\pi$-exchange NN potential generated at N$^2$LO constitutes a chiral EFT 
prediction, which is well supported phenomenologically. In particular, clear
evidence of the parameter-free $2\pi$-exchange
potential at orders N$^2$LO and N$^4$LO was observed by analyzing
neutron-proton and proton-proton scattering data in
Refs.~\cite{Epelbaum:2014efa,Epelbaum:2014sza,Reinert:2017usi}, see
also Ref.~\cite{Epelbaum:2024gfg} for a related discussion.  Moreover, despite
the already mentioned enhancements, a regular convergence
pattern of chiral EFT in the NN sector was confirmed for the
SMS potentials of Ref.~\cite{Millican:2025sdp} with $\Lambda =
450$ and $500$~MeV, assuming the breakdown scale of $\Lambda_b =
600$-$700$~MeV. These results in the two-nucleon sector suggest that
the enhanced N$^4$LO 
3NF contributions are not necessarily indicative of a slow
convergence of chiral EFT, which can only be analyzed quantitatively after taking
into account all contributions up to the considered order and performing implicit 
renormalization by fixing the LECs of the short-range interactions
from experimental data. It is also worth mentioning that the
enhancement of the subleading two-pion exchange potentials due to the
large values of the pion-nucleon LECs $c_i$ can be mitigated by using
the $\Delta$-full formulation of chiral EFT, see
Refs.~\cite{Ordonez:1995rz,Kaiser:1998wa,Krebs:2007rh,Epelbaum:2008td}
and \cite{Epelbaum:2007sq,Krebs:2018jkc}
for applications to the NN and 3N potentials. In this
  framework, a part of the N$^4$LO (N$^5$LO) 3NF contributions driven
  by the LECs 
  $c_{2,3,4}$ ($F_2$) is shifted to N$^3$LO.
\end{itemize}
Clearly, the qualitative conclusions and estimates made in this paper should be
backed by explicit calculations including higher-order corrections
to the 3NF. Work along these lines is in progress by the LENPIC
Collaboration using the recently proposed gradient flow regularization
method \cite{Krebs:2023ljo,Krebs:2023gge}.

\acknowledgments
We appreciate useful discussions with Vincenzo Cirigliano, Wouter
Dekens, Sanjay Reddy, Bira van Kolck and Kai Hebeler. We are also grateful to the
authors of Ref.~\cite{Cirigliano:2024ocg} for providing their internal
notes on the estimation of the impact of the N$^2$LO 3NF in dense
matter. 
Two of the authors (EE and HK) would like to thank the local organizers of
the workshop ``Standard Model EFT meets Chiral EFT (SMEFT meets
ChEFT)'', held in Vancouver during September 29 -- October 3, 2025, for their
hospitality and for providing a forum to discuss our results with the
authors of Ref.~\cite{Cirigliano:2024ocg}.
This work has been supported by the European Research Council (ERC)
under the European Union’s Horizon 2020 research and innovation
programme (grant agreement No.~885150), by the MKW NRW under the
funding code NW21-024-A, by JST ERATO (Grant No. JPMJER2304), by JSPS
KAKENHI (Grant No. JP20H05636) and by the Georgian Shota Rustaveli National 
Science Foundation (Grant No. FR-23-856).

\bibliography{6.0}
\bibliographystyle{apsrev}

\end{document}